\documentclass[
	twocolumn,
	superscriptaddress,
	preprintnumbers,
	amsmath,
	amssymb,
	floatfix,
	nofootinbib
]{revtex4-1}

\usepackage{graphicx}
\usepackage{blkarray}
\usepackage[group-separator={,}]{siunitx}
\usepackage{tikz-cd}

\begin{document}

\title{Dynamical phases in growing populations: understanding recovery from bottlenecks}

\author{Emanuele Crosato}
\thanks{These authors contributed equally to the manuscript.}
\affiliation{School of Physics and EMBL Australia node in Single Molecule Science, School of Medicine, UNSW, Sydney, Australia}	
 
\author{Jeffrey N.~Philippson}
\thanks{These authors contributed equally to the manuscript.}
\affiliation{Simons Centre for the Study of Living Machines, National Centre for Biological
Sciences, Tata Institute of Fundamental Research, GKVK Campus, Bellary Road, Bengaluru 560065, India}

\author{Shashi Thutupalli}
\thanks{To whom correspondence should be addressed: \texttt{shashi@ncbs.res.in} and \texttt{r.g.morris@unsw.edu.au}.}
\affiliation{Simons Centre for the Study of Living Machines, National Centre for Biological
Sciences, Tata Institute of Fundamental Research, GKVK Campus, Bellary Road, Bengaluru 560065, India}
\affiliation{International Centre for Theoretical Sciences, Tata Institute of Fundamental Research, Survey no 151, Shivakote, Hesaraghatta Hobli, Bengaluru 560089, India}

\author{Richard G.~Morris}
\thanks{To whom correspondence should be addressed: \texttt{shashi@ncbs.res.in} and \texttt{r.g.morris@unsw.edu.au}.}
\affiliation{School of Physics and EMBL Australia node in Single Molecule Science, School of Medicine, UNSW, Sydney, Australia}
\affiliation{Simons Centre for the Study of Living Machines, National Centre for Biological
Sciences, Tata Institute of Fundamental Research, GKVK Campus, Bellary Road, Bengaluru 560065, India}

\begin{abstract}
	Since steep declines in a population's size also typically alter its composition, population bottlenecks are considered highly important for evolution. However, despite such significance, the mechanisms governing the impact of a given population bottleneck remain poorly understood. In this context, we show that long-term post-bottleneck outcomes can depend crucially on the rate at which a diminished population grows whilst recovering. That is, two otherwise identical populations, each having undergone the same bottleneck, can fixate on dramatically different demographics due to different rates of post-bottleneck growth. This behaviour is moreover shown to change non-trivially with different levels of mutation. The underlying mechanism can be traced to intrinsic fluctuations whose standard deviation decreases in time with the inverse square root of the growing population size. Crucially, such decreasing fluctuations couple to the underlying dynamics and result in distinct regimes of (non-equilibrium) demographic behaviour, delimited by abrupt transitions at critical population sizes. Seen through this lens, growth and mutation alter evolutionary outcomes by changing the duration and character of dynamical phases; a feature that we speculate may be of generic importance across many systems.
	%
\end{abstract}

\maketitle

\section{Introduction}

Population bottlenecks involve a steep decline in population size due to exogenous events, such as disease, changes to the climate, or population fracture (often referred-to as a founder event) \cite{mayr_change_1954,Nei1975}. They are widely accepted as an important facet in the modern understanding of evolution, and have been implicated in the reduction of both genetic and phenotypic variation across a variety of organisms, including viruses \cite{domingo2012viral, weaver2021population}, song sparrows \cite{Keller2001,Keller1994a}, tropical surgeonfish \cite{Doherty2004}, elephant seals \cite{Hoelzel2002a} and humans \cite{Manica2007a}.

Despite such conceptual importance, the relationship between the short-term restriction of a population and its subsequent long-term evolutionary outcomes is still poorly understood, with a lack of both representative models and/or classifying phenomenology.
This is particularly relevant since post-bottleneck recovery is typically characterised by rapid or sustained population growth, which several pioneering works have now shown can lead to a variety of ostensibly counter-intuitive behaviours. These include driving novel demographic transients \cite{Melbinger2010}; facilitating cooperation \cite{Cremer2011, Huang2015}, even when deterministic selection favours defection \cite{Constable2016}; and altering the success of invading variants \cite{Ashcroft2017}.

We therefore set out to disentangle the effects of a bottleneck itself---manifest as the population's initial size and demographic mix following the bottleneck---and those of post-bottleneck characteristics, such as the rates of growth and mutation. To do this, we use evolutionary game theory \cite{smith1982evolution, hofbauer1998evolutionary, szabo2007evolutionary} and, specifically, a growing variant of the otherwise well-studied Iterated Prisoner's Dilemma (IPD) under replicator-mutator dynamics \cite{Imhof2005, Bladon2010}.

The growing IPD has several appealing characteristics; it is at once minimal, facilitating both analysis and simulation, but it is also sufficiently rich as to act as an effective `toy' system for understanding broader phenomenology. For example, the IPD includes two stable attractors, one of which is a limit cycle, which speaks to models of stable population diversity, such as that observed in \textit{Escherichia coli} \cite{durrett1997allelopathy, kerr2002local, kirkup2004antibiotic} and side-blotched lizards \cite{sinervo1996rock} that are understood via cyclic dominance \cite{szolnoki2014cyclic}. The IPD also crudely captures the effects of mutation, which have so far been overlooked in the literature concerning growth, but are shown here to be crucial.

Ultimately, we demonstrate that the post-bottleneck rates of population growth and/or mutation critically determine how fixation probabilities are conditioned on initial population size and demographic mix. This means that even marginal differences to the rates of growth or mutation can result in dramatically different likelihoods of long-term outcomes, despite starting from the same founding population. For example, a bottleneck that restricts the population's composition to the location of one of the two stable attractors typically results in long-term fixation to the same attractor. However, for certain combinations of the rates growth and mutation, this is turned on its head, and long-term fixation to the other attractor occurs with almost certainty. This counter-intuitive and critical-like behaviour broadly interpolates between known behaviours in the deterministic, large population limit, where fixation is fully determined by initial conditions, and systems at finite population size, that cannot fixate due to persistent fluctuations. 

The underlying mechanism, generic across well-mixed populations, is an intrinsic noise whose standard deviation \emph{decreases} in time with the inverse square root of the growing population size, $N$. This ensures that, as a population grows, its dynamics will eventually fixate upon the stable attractors that characterise the system in the limit of very large population size. However, it also permits a diverging correlation time, so that long-term outcomes can depend on early events, even as $t\to\infty$, despite the ostensibly randomising effects of intervening fluctuations \cite{Klymko2017}. Moreover, the \emph{rate} at which fluctuations are suppressed turns out to be vital to understanding the statistics of long-term fixation.

Crucially, the decreasing scale of the fluctuations couples to the geometry of the deterministic attractors, manifesting in distinct regimes of demographic behaviour. These can be explained in terms of \emph{effective} non-equilibrium phases, which capture the transient characteristics of fixed-$N$ systems by averaging over large but finite times (hence excluding the disproportionate effect of certain highly rare events). Such phases are moreover shown to be delimited by transitions at critical values of $N$. This is reminiscent of the classical study of non-equilibrium phase transitions \cite{Hinrichsen2000a, Odor2004a}, which have previously been implicated across a number of areas, including directed percolation \cite{Broadbent1957}, self-organisation of particle suspensions \cite{Corte2008}, surface growth \cite{Tu1997a}, epidemiology \cite{Cardy1985a} and even hard sphere packing \cite{Wilken2021}. It is within this framework that we understand the importance of the rates of population growth and mutation, which together alter the character and duration of such regimes, and hence determine the long-term impact of bottlenecks.

Our results are organised as follows. In Section \ref{sec:GIPD} we introduce the growing IPD as a toy model for understanding recovery from bottlenecks. We demonstrate that there are only three possible outcomes in the long-time limit: fixation on one of the two attractors or extinction. In Section \ref{sec:stats} we compute the likelihood of such fixation as a function of bottleneck (initial conditions), and the rates of growth and mutation. This illustrates our first key message: growth and mutation can critically dictate the impact of a given bottleneck on evolutionary outcomes (long-term fixation probabilities). To understand these results, Section \ref{sec:phases} outlines our second key message: dynamical behaviour can always be characterised by one of three effective non-equilibrium phases, dependent on the population size. These are: 
\begin{enumerate}
	\item a \emph{stochastically-induced} phase, at small population sizes, where state-dependent fluctuations dominate, and crossings between both basins of attraction are permitted; 
	\item an \emph{asymmetric} phase at intermediate population sizes, where crossings between the two basins of attraction are overwhelmingly likely to be in one direction only, and; 
	\item a \emph{locked-in} phase, at large population sizes, where escape from the basins of attraction is extremely unlikely.
\end{enumerate}
Section \ref{sec:growthandmutation} then shows how a growing population exhibits these phases sequentially, in a manner controlled by the rates of growth and mutation. This suggests that our first message can be explained in terms of the second. Finally, Section \ref{sec:equiv} validates this conjecture by satisfactorily reconstructing the outcome statistics of Section \ref{sec:stats} using a decomposition of conditional probabilities motivated by the observed dynamical phases.
\begin{figure}[t!]
\centering
\includegraphics[width=0.99\columnwidth]{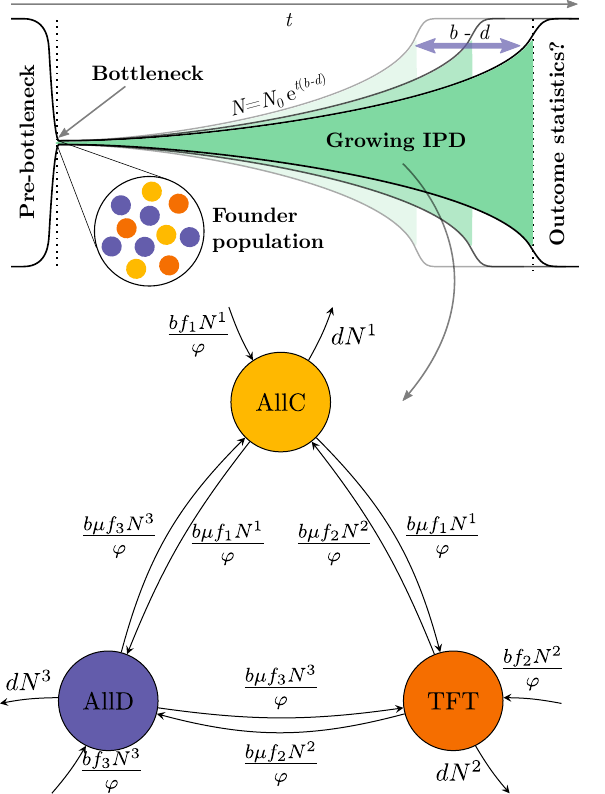}
\caption{
	\textbf{The growing frequency-dependent IPD: a toy model for bottleneck recovery via population growth}.
By de-coupling the rates of birth and death in the frequency-dependent IPD \cite{Imhof2005}, we adopt a model
that combines the notions of frequency-dependence and mutation with that of population growth (see main text for definitions of rates). We ask: what is the long-term effect of a population bottleneck?  Specifically, how does it influence evolutionary outcomes, and how does this change with the rate at which the recovering population grows?
}
\label{fig:setup}
\end{figure}

\section{Results}

\subsection{Growing frequency-dependent IPD}\label{sec:GIPD}

\begin{figure*}[t]
\centering
\includegraphics[width=\textwidth]{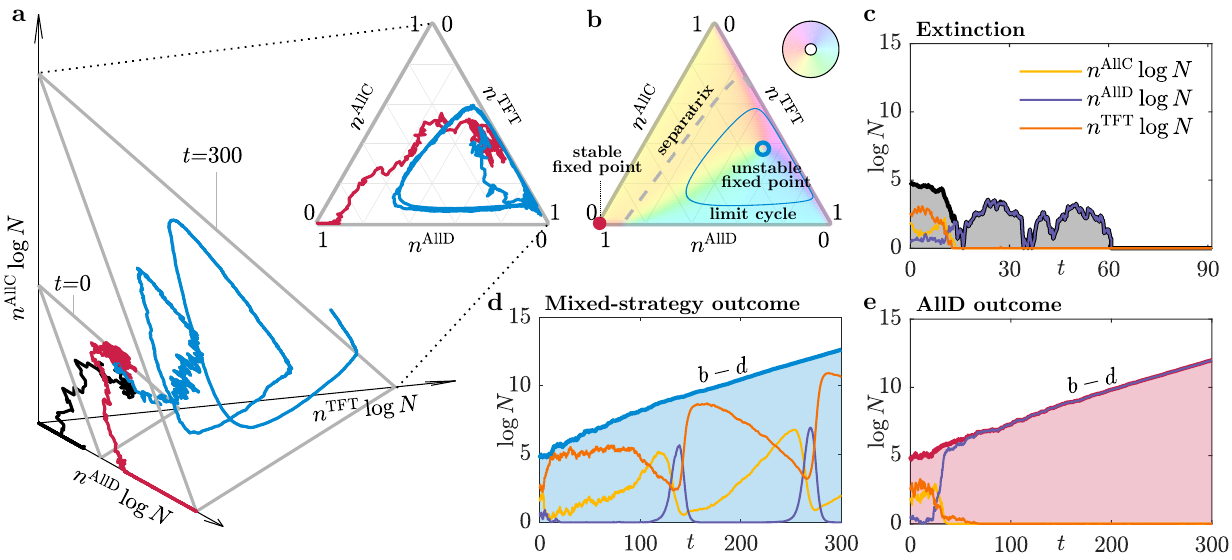}
\caption{
	\textbf{Barring extinction, recovering populations fixate on one of two deterministic attractors}. Projecting the dynamics of the growing IPD (panel \textbf{a}) onto the unit simplex (panel \textbf{a}, inset) demonstrates $O(1/\sqrt{N})$ intrinsic fluctuations that decrease as the population grows in time. Since overall population growth is exponential (with rate $b-d$) the system converges on canonical deterministic dynamics as $t\,{\to}\,\infty$ (panel {\bf b}, colours represent the direction of deterministic flows). This results in only three different evolutionary outcomes.  Either, the population goes extinct (panel \textbf{a} in black, panel \textbf{c}), or the system fixates on one of the two deterministic attractors: the mixed-strategy limit cycle (panels \textbf{a} in blue, panel \textbf{d}) or the AllD fixed point (panels \textbf{a} in red, panel \textbf{e}). The statistics of these two latter outcomes are dictated not only by the founding population, but by fluctuations. These determine the likelihood of crossing the separatrix that marks the boundary between the two basins of the stable attractors (dashed, panel \textbf{b}). Lower-case font represents population fractions, \textit{e.g.}, $n^\mathrm{AllD} = N^\mathrm{AllD}/N$.
}
\label{fig:1}
\end{figure*}
\begin{figure*}[t]
\centering
\includegraphics[width=\textwidth]{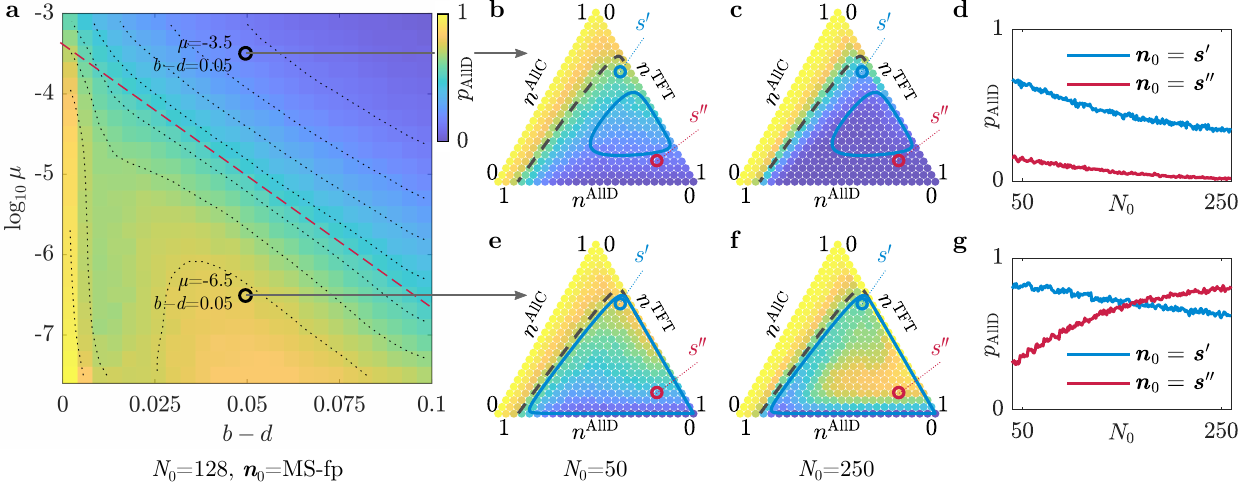}
\caption{
	\textbf{Rates of growth and mutation critically alter how a given bottleneck translates into fixation probabilities}. For a range of initial conditions, growth and mutation rates, we use repeated Gillespie-It\^{o} simulations to compute $p_\mathrm{AllD}$, the probability that a growing population will fixate on the AllD fixed point.  Choosing the initial demographic mix as the mixed-strategy fixed point, we see that $p_\mathrm{AllD}$ depends on both the rate of population growth and the rate of mutation (panel {\bf a}).  Conversely, fixing the rates of growth and mutation, $p_\mathrm{AllD}$ also depends on both the initial demographic mix $\boldsymbol{n}_0$ (panels \textbf{b}, \textbf{c}, \textbf{e} \& \textbf{f}) and the initial population size, $N_0$ (panel \textbf{d} \& \textbf{g}). Behaviour can be qualitatively classified into two regions separated by the red dashed line (panel {\bf a}).  In the upper region, behaviour agrees with expectations: higher rates of growth more rapidly suppress $O(1/\sqrt{N})$ intrinsic fluctuations and therefore increase the likelihood of fixation within the basin of attraction that the system started (panel {\bf a}).  Increasing the initial population size reduces the likelihood of large fluctuations at early times, therefore exacerbating this effect (panels {\bf b}-{\bf d}).  In the lower region, behaviour is more complex: there are high likelihoods of an AllD outcome at both low ($b\,{-}\,d\,{\lesssim}\,0.01$) and intermediate ($0.03\,{\lesssim}\,b\,{-}\,d\,{\lesssim}\,0.07$) rates of growth (panel {\bf a}).  This results in a non-monotonic dependence of $p_\text{AllD}$ on $b-d$, which implies that population growth can actually {\it increase} the likelihood of crossing the separatrix between the two basins of attraction, despite more rapidly suppressing $O(1/\sqrt{N})$ fluctuations.  There is a similarly non-trivial structure to the effects of initial demographic mix on $p_\text{AllD}$ where, for certain $\boldsymbol{n}_0$, an increase in $N_0$ actually causes $p_\text{AllD}$ to increase rather than decrease (representative initial states $\boldsymbol{s}^{\prime\prime}$ and $\boldsymbol{s}^\prime$, respectively, panels {\bf e}-{\bf g}).  
}
\label{fig:2}
\end{figure*}

The frequency-dependent IPD is a well-established evolutionary game involving three strategies.
It abstracts the key facets of biological evolution and population dynamics (e.g., birth, death, and mutation) into a tractable framework whereby the fitness of individuals is determined by their relative successes and/or failures when playing each other in the repeated game, the IPD.

In the classical formulation, each player of the repeated IPD assumes one of three strategies: ``always cooperate'' (AllC), who cooperate in every round; ``always defect'' (AllD), who defect in every round, or; ``tit-for-tat'' (TFT), who default to cooperation for the first round and then, at a small complexity cost, copy their opponents' moves thereafter.
In a single repeated game, players accumulate payoff over the rounds according to the standard Prisoner's Dilemma rules: if both players cooperate, they receive a larger pay-off than if they both defect, but if one player cooperates and the other defects, then the defector receives the highest possible pay-off whilst the cooperator gets the lowest payoff.
The accumulated payoffs for an $m$-round repeated game are encoded by the matrix \cite{Imhof2005}:
\begin{equation}
\nonumber
	\begin{blockarray}{lccc}
	\quad & \text{AllC} & \text{AllD} & \text{TFT} \\
	\begin{block}{l(ccc)}
	\text{AllC \ } & Rm & Sm & Rm \\
	\text{AllD \ } & Tm & Pm & T + P(m-1)\\
	\text{TFT \ } & Rm - c & S + P(m-1) -c & Rm - c \\
	\end{block}
	\end{blockarray}
\end{equation}
where $T>R>P>S$ and $R>(T+S)/2$. We use $T=5$, $R=3$, $P=1$, $S=0.1$, and $c=0.2$.

This game is then set against a backdrop of birth, death and mutation (Fig.~\ref{fig:setup}). With a rate $b$, the fraction of players born into a given strategy is proportional to the fraction of the total pay-off accumulated by that strategy when all players play each other. This is given by $f_i N^i / \varphi$, where 
\begin{equation}
	f_i = \frac{\sum_{j=1}^3 a_{ij} N^j - a_{ii}}{N-1},
	\label{eq:fit}
\end{equation}
is the average fitness of a given strategy when played against the whole population (including individuals of the same strategy). Here, Latin indices denote the three different strategies, i.e., $\textrm{AllC}\,{\rightarrow}\,1$, $\textrm{AllD}\,{\rightarrow}\,2$ and $\textrm{TFT}\,{\rightarrow}\,3$ such that the $a_{ij}$ are the components of the above payoff matrix and $N^i$ is the number of individuals playing each strategy.  In our analysis, but not in simulations, we use the simplification that
\begin{equation}
	f_i = \frac{\sum_{j=1}^3 a_{ij} N^j}{N},
	\label{eq:fit_2}
\end{equation}
which, although including self-interactions, still retains all the relevant features associated with the IPD \cite{Bladon2010}.  In either case, the mean fitness is $\varphi = \sum_{i=1}^3 f_i\,N^i/N$. As a result, the higher the relative fitness associated with a strategy, the more likely that individuals are born with that strategy. A small fraction $\mu$ of births further undergo mutation and are assigned a different strategy.  Death also occurs at random, with a rate $d$.

Choosing a population birth rate $b$ that is greater than the death rate $d$ gives rise to unbounded exponential ({\it i.e.}, Malthusian) population growth (Fig.~\ref{fig:1}{\bf a}).  This rapidly suppresses $O(1/\sqrt{N})$ fluctuations and converges to well-studied deterministic behaviour (represented on the unit simplex in Fig.~\ref{fig:1}{\bf a}, inset). In particular, for values of mutation rate $\mu$ in the interval $10^{-7.5}\,{\leq}\,\mu\,{\leq}\,10^{-2.5}$, there are two stable deterministic attractors \cite{Imhof2005}: a stable `AllD' fixed point, where a small fraction of TFT players (who mutually cooperate) are exploited by a large population of AllD defectors, and; a stable limit cycle around an unstable `mixed-strategy' fixed point (Fig.~\ref{fig:1}{\bf b} \& Supplementary Information, Sec.~\ref{sec:det-lim}).  The latter is characterised by a three-phase cycle whose handedness is anti-clockwise in the traditional presentation of the state-space simplex; players of TFT can out-compete those playing AllD due to their capacity for mutual cooperation, however: they are then susceptible to invasion by players of AllC due to the complexity cost, whereby; AllC players can be exploited by those playing AllD, completing the cycle.

As a consequence, only three outcomes are possible as $t\,{\to}\,\infty$.  Either \textit{i}) the population goes extinct in the early stages due to finite size fluctuations (Figs.~\ref{fig:1}{\bf a}-black \& {\bf c}) or, its demographic mix converges on \textit{ii}) the mixed-strategy limit cycle (Figs.~\ref{fig:1}{\bf a}-blue \& {\bf d}) or \textit{iii}) the AllD fixed point (Figs.~\ref{fig:1}{\bf a}-red \& {\bf e}).

\subsection{Statistics of evolutionary outcomes}\label{sec:stats}

Since the growing IPD is both non-linear and time-inhomogeneous, it resists most standard approaches to probabilistic analysis. Computing the statistics of the aforementioned $t\to\infty$ outcomes---{\it i.e.}, the fixation probabilities---therefore involves using a high performance computing facility \cite{Smith_Betbeder_Matibet_2010} to perform stochastic simulations.

Specifically, we employ a hybrid Gillespie-It\^{o} approach (see Supplementary Information, Sec.~\ref{sec:simulations}) to approximate the following fixation probability:
\begin{equation}
	p_\mathrm{AllD} = \lim_{t\to\infty}\Pr\big\{\boldsymbol{n}_t=\text{AllD-fp}\,\vert\,\boldsymbol{N}_0,\,N_{t^\prime}> 0\ \forall\ t^\prime \leq t\big\},
\label{eq:palld}
\end{equation}
where $\boldsymbol{n}_t\,N_t = \boldsymbol{N}_t = \{N_t^\text{AllC}, N_t^\text{AllD}, N_t^\text{TFT}\}$. This is the limiting probability, after long times, that the system converges to the AllD fixed point, given specified post-bottleneck initial conditions, and conditioned on populations that do not become extinct.

The results demonstrate several interesting features. For instance, choosing $\boldsymbol{n}_0$ to be the unstable mixed-strategy fixed point, we see that $p_\text{AllD}$ depends on both the rate of population growth and the rate of mutation (Fig.~\ref{fig:2}{\bf a}).  Similarly, fixing the rates of population growth and mutation reveals sensitivity to initial conditions, where $p_\text{AllD}$ depends on both initial demographic mix, $\boldsymbol{n}_0$, and initial population size, $N_0$ (Figs.~\ref{fig:2}{\bf b}-{\bf g}).

The parameter space of mutation and growth rates can moreover be divided into two qualitative regions (panel {\bf a}, red dashed line).

In the upper region, behaviour agrees with the expectation that faster growth rates reduce $O(1/\sqrt{N})$ intrinsic fluctuations more rapidly and therefore increase the likelihood of fixation within the same basin of attraction that the system started (panel {\bf a}). Similarly, as the initial population size, $N_0$, increases, the system experiences fewer large fluctuations at early times and hence this also increases the likelihood of fixation within the starting basin of attraction (Figs.~\ref{fig:2}{\bf b}-{\bf d}).

In the lower region, however, behaviour is more complex. There are high likelihoods of an AllD outcome at both low ($b\,{-}\,d\,{\lesssim}\,0.01$) and intermediate ($0.03\,{\lesssim}\,b\,{-}\,d\,{\lesssim}\,0.07$) rates of growth (panel {\bf a}).  This results in a non-monotonic dependence of $p_\text{AllD}$ as a function of population growth. That is, despite more rapidly reducing the fluctuations that are ostensibly required to cross the separatrix between mixed-strategy and AllD basins, growth can actually {\it increase} the likelihood of fixating on the AllD fixed point.  Similarly, rather than decreasing the likelihood of crossing the separatrix and fixating there, increasing $N_0$ can actually increase this likelihood for certain initial states, confounding expectations regarding the role of fluctuations at early times when populations remain small (Figs.~\ref{fig:2}{\bf e}-{\bf g}).

In the context of population bottlenecks, this demonstrates that the long-term ramifications of reducing a population to a particular size and demographic mix can depend, critically, on the rates of post-bottleneck growth and mutation.

\subsection{Effective non-equilibrium phases}\label{sec:phases}

\begin{figure*}[t]
\centering
\includegraphics[width=\textwidth]{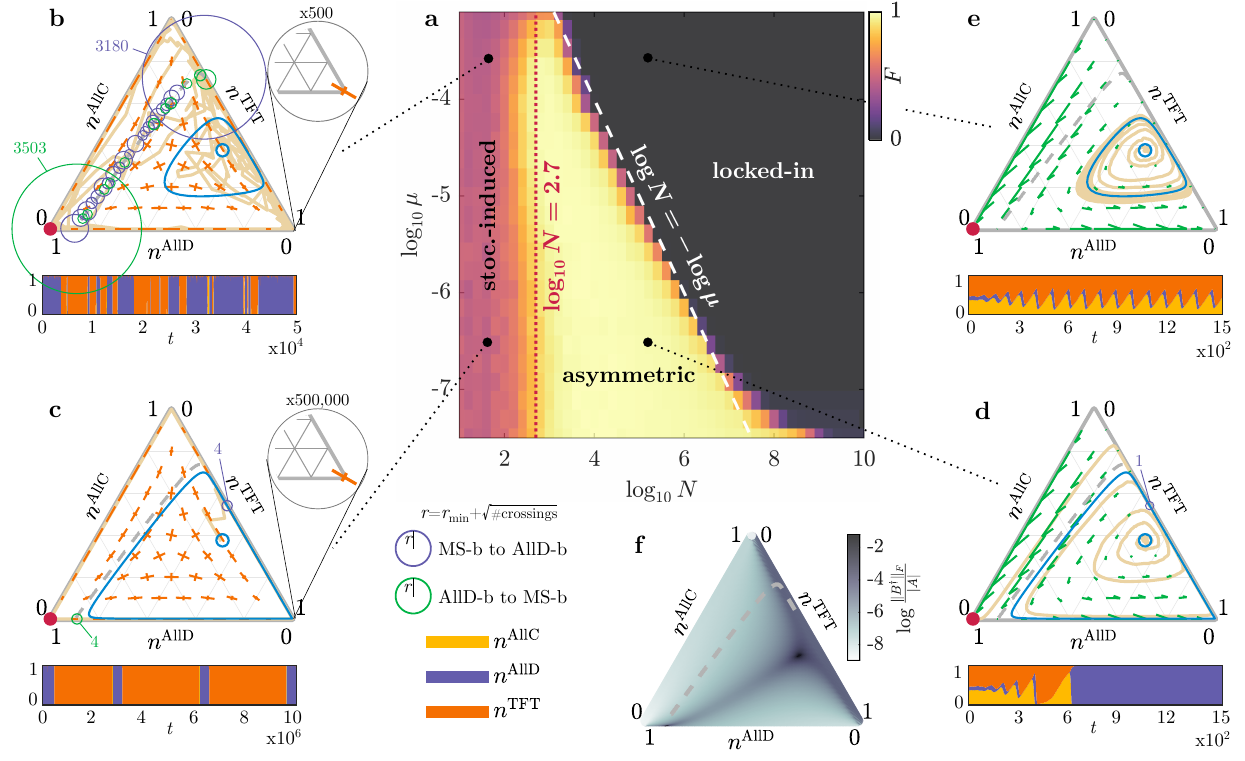}
\caption{
	\textbf{Effective non-equilibrium phases}. When characterised by the fraction $F$ (of a large but finite time) that trajectories spend in the basin of the AllD fixed point, repeated simulations of the IPD at fixed population sizes demonstrate three distinct regimes of demographic behaviour (panel \textbf{a}). When population sizes are small, fluctuations are dominant, and the separatrix can be crossed in both directions (circles, panels \textbf{b} \& \textbf{c}). Eigenvalues of the projected correlation matrix, $B^\dagger_{ij}$ indicate that the magnitude and bias of the fluctuations depend on demographic mix (orange crosses, panels \textbf{b} \& \textbf{c}). Generally, this gives rise to fluctuation gradients that drive the system towards the simplex edges and corners. Although, due to the structure of the payoff matrix there is also an anti-clockwise bias and a lower probability of finding the AllC corner. At the simplex boundaries, normal fluctuations are proportional to $\sqrt{\mu}$ (panels \textbf{b} \& \textbf{c}, magnified inset). Characteristic trajectories (beige) are therefore increasingly confined to the boundaries and corners as $\mu$ decreases. Since the residence times associated with the corners are $O(1/\mu)$, trajectories spend a disproportionate fraction of their time in the AllD and TFT corners (bars, panels \textbf{b} \& \textbf{c}).  Beyond this regime, behaviour becomes increasingly deterministic (green half-arrows, panels \textbf{e}-\textbf{d}). For intermediate population sizes, behaviour is \textit{asymmetric}: demographic trajectories can cross the separatrix from the mixed-strategy limit cycle, but not from the AllD fixed point (panel \textbf{d}). The upper critical population size of this regime scales as $\sim 1/\mu$ (white, panel \textbf{a}). For large enough $N$, behaviour becomes increasingly deterministic, and trajectories remain locked-in to the mixed-strategy limit cycle (beige, panel \textbf{e}).  At low $\mu$, the separatrix is typically crossed where it intersects either the TFT or AllD edges, which is where the magnitude of stochastic effects, $\Vert B^\dagger\Vert_F$, are largest relative to deterministic flows, $\vert A_i\vert$, with $\Vert\cdot\Vert_F$ and $\vert\cdot\vert$ denoting Frobenius- and $\ell^2$-norms, respectively (panels {\bf c}, {\bf d} \& {\bf f}).  Representative trajectories and population fractions (bars) only show a fraction of the total time simulated (see bar legend) in order to aid visualisation. Crossing statistics (circles) are taken from single simulations lasting $10^6\,$s.
}
\label{fig:3}
\end{figure*}
To understand the non-trivial behaviour in Fig.~\ref{fig:2}, we repeatedly simulate the \textit{fixed} population size IPD (Supplementary Information, Secs.~\ref{sec:simulations} \& \ref{sec:lan-eq-fix-size}), computing a so-called empirical distribution~\cite{touchette_introduction_2018}. Specifically, we calculate the mean fraction of time spent in the AllD basin,
\begin{equation}
	F=\left\langle\frac{1}{T}\int_0^T 1_{\boldsymbol{n}_t\in\text{AllD-b}}\,dt\right\rangle_\mathcal{N},
\end{equation}
where $1$ is the indicator function, $\mathcal{N}$ denotes the size of the ensemble over which the average is taken, and the integration is understood in the It\^{o} sense. The time $T>\left(\log N_\mathrm{max}\right) / (b-d)$ exceeds the entire duration of our growing simulations (which we stop at $N_\mathrm{max}=10^{10}$) and represents an effective cutoff, so that the statistics of fixed size simulations are not skewed by events that are highly unlikely to occur in the growing simulations ({\it i.e.}, with characteristic rates $\ll1/T$). Such a large but finite $T$ therefore aims to capture the average transient behaviour of a growing population at a particular $N$.

\begin{figure*}
\centering
\includegraphics[width=\textwidth]{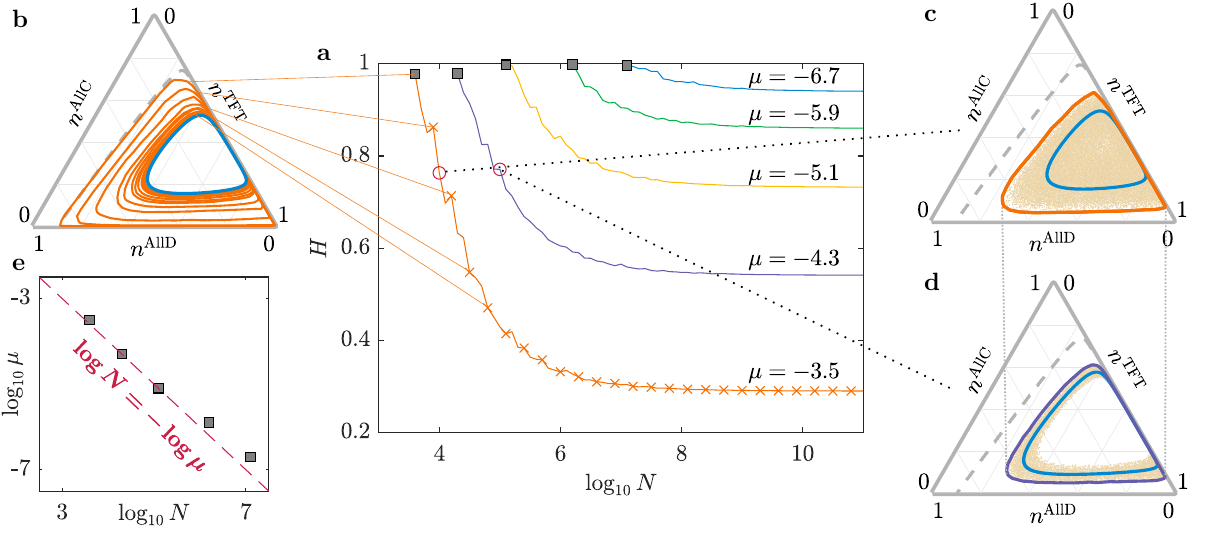}
\caption{
	\textbf{Understanding the mutation-dependent transition between asymmetric and locked-in phases}. At finite population sizes (and for times $t<T$) demographic trajectories in the locked-in regime repeatedly circumnavigate the limit cycle, leading to a well-defined `footprint' of visited points due to the state-dependence of the noise correlations.  We compare such footprints via the convex hull, $\text{Hull}\left(\boldsymbol{n}_{t<T}\right)$, of all points sampled by a trajectory $\boldsymbol{n}_{t<T}$ within a specified (large) time, $T$ (panel {\bf a}). The area of this convex hull is written as a fraction of the area $A$ of the mixed-strategy basin of attraction, $H = \text{Area}\left[\text{Hull}\left(\boldsymbol{n}_{t<T}\right)\right]/A$. For a given $\mu$, the fraction $H$ converges exponentially to the value associated with the deterministic limit cycle, as $N$ increases (panels {\bf a} \& {\bf b}).  As a result, there are approximate equivalences between the footprint of trajectories with a large $\mu$ and small $N$ and those with a correspondingly smaller $\mu$ and larger $N$ (panels {\bf a}, \textbf{c} \& \textbf{d}). This leads to a family of scaling relations in the locked-in regime (panel {\bf a}). The {\it critical} scaling, at which the footprint occupies the entire mixed-strategy basin, is $N\sim1/\mu$, which agrees with ensemble statistics (panel {\bf e}).  Since this corresponds to onset of the locked-in phase, the corollary is that the asymmetric phase has a duration that is $\mu$-dependent ({\it cf.}~Fig.~\ref{fig:3}). 
}
\label{fig:4}
\end{figure*}
Our results (Fig.~\ref{fig:3}\textbf{a}) are suggestive of a large deviation principle, such that $p(F\,\vert\,\mathcal{N})\asymp \exp{\left[-\mathcal{N}\,I_\mu\left(F_\mathcal{N}\right)\right]}$, where $I_\mu$ is a convex rate function.  Although determining the precise functional form of $I_\mu$ is considered out-of-scope for this article, our data suggests that it has only three zeros, despite varying $N$ over ten orders of magnitude:
\begin{equation}
	\text{arg\,min}\,I_\mu \left(F\right) \approx \left\{
		\begin{aligned}
			&0.5,\ \ \forall\ N\leq 10^{2.7} \\
			&0,\ \ \forall\ 10^{2.7}<N<1/\mu \\
			&1,\ \ \forall\ N>1/\mu
		\end{aligned}\right.\quad .
\end{equation}
This means that, depending on the population size, there are three statistically distinct types of characteristic demographic behaviour. Due to the finite size of $T$, we call these \emph{effective} non-equilibrium phases. The three effective phases are characterised as follows: 
\begin{enumerate}
	\item \textit{Stochastically-induced phase}--- Demographic trajectories at small population sizes are characterised by large intrinsic fluctuations and an intermediate value of $F$ (Fig.~\ref{fig:3}\textbf{a} in magenta, Figs.~\ref{fig:3}\textbf{b} \& \textbf{c}). Fluctuations are both correlated and state-dependent; features that are captured by the symmetric $3\times 3$ correlation matrix, $B_{ij}^\dagger$, that can be obtained by performing a Van Kampen system-size expansion \cite{van_kampen} and projecting the results onto the unit simplex using It\^{o}'s lemma (Fig.~\ref{fig:3}\textbf{b} \& \textbf{c}, orange crosses, and Supplementary Information, Sec.~\ref{sec:proj}).  The frequency-dependent nature of births means that fluctuations at the centre of the simplex are large and isotropic, whilst the components normal to the boundaries decrease rapidly as the edges and corners are approached.  This gives rise to stochastically-induced effects \cite{Horsethemke2006,Jhawar2020}, where fluctuation gradients bias stochastic trajectories, driving them towards the simplex edges and corners, on average (Fig.~\ref{fig:3}\textbf{b} \& \textbf{c} and Supplementary Information Secs.~\ref{sec:sse}-\ref{sec:proj}). Despite such overall behaviour, $B_{ij}^\dagger$ is not symmetric under the interchange of $n^\text{AllC}$, $n^\text{AllD}$ and $n^\text{TFT}$, and stochastic trajectories retain characteristics encoded by the payoff matrix; including a bias for anti-clockwise motion, and a comparatively low likelihood of reaching the AllC corner (when compared to AllD and TFT corners).  These behaviours are crucial to understanding the precise $\mu$-dependent mechanisms that underpin the value $F\approx 0.5$. 
	\item \textit{Asymmetric phase}--- Increasing $N$, the magnitude of fluctuations decreases, and the relative geometry of the underlying attractors becomes increasingly important.  In particular, the system enters an asymmetric regime at populations above $N\approx 10^{2.7}$, for which $F\approx 1$ (Fig.~\ref{fig:3}\textbf{a}, yellow). Here, state-dependent fluctuations permit the system to cross from the mixed-strategy limit cycle to the AllD basin, but not from the AllD fixed point to the basin of the limit cycle (Fig.~\ref{fig:3}\textbf{d}).  In other words, once the separatrix has been crossed, trajectories are extremely unlikely to come back within the time $T$.
	\item \textit{Locked-in phase}--- Once $N$ is sufficiently large, fluctuations are small and demographic trajectories are effectively locked into the mixed-strategy basin for times $<T$, implying $F\approx 0$ (Fig.~\ref{fig:3}\textbf{e}, dark gray).
\end{enumerate}

The aforementioned effective phases moreover couple to mutation, which alters both the character of the stochastically-induced phase (Figs.~\ref{fig:3}\textbf{b} \& \textbf{c}), and the population size at which the system transitions from asymmetric to locked-in phases (Fig.~\ref{fig:3}\textbf{a}, white dashed line).

The origin of the former is that the standard deviation of fluctuations normal to the boundaries is $O(\sqrt{\mu}/N)$ (Supplementary Information, Sec.\ref{sec:boundary}).  Evolutionary trajectories therefore become increasingly confined to the boundaries as $\mu$ decreases (Figs.~\ref{fig:3}\textbf{b} \& \textbf{c}). This not only exacerbates stochastically-induced effects, but also increases the mean residence times associated with the corners. In particular, at comparatively high levels of mutation, residence times are less than $T$, which results in a stochastic cycling between TFT and AllD corners (recall that there is a lower likelihood of finding the AllC corner) (Fig.~\ref{fig:3}\textbf{b}). Since these two corners have comparable mean rates of escape to the opposite basin of attraction (Supplementary Information, Sec.\ref{sec:escape-time}), $F$ takes a value of approximately $0.5$.  By contrast, lower rates of mutation imply dwell times greater than $T$ (Fig.~\ref{fig:3}\textbf{c}).  On average, therefore, the system will find either the AllD or TFT corner and then remain there.  Here, the value of $\approx 0.5$ results from the position of the mixed strategy fixed point, and the correspondingly equal probability that trajectories are expelled to either the AllD or TFT corners (typically, via the AllC-TFT edge, see Fig.~\ref{fig:3}\textbf{f}).


For the latter, the $\sim 1/\mu$ dependence of the asymmetric to locked-in transition can be understood in terms of the stochastic `footprint' of evolutionary trajectories in the locked-in regime--- {\it i.e.}, those that repeatedly (and stochastically) navigate the limit cycle (Fig.~\ref{fig:4}). In particular, due to the finite nature of $T$, the convex hull of this footprint is well-defined, reflecting the shape of the limit cycle at a {\it different} value of $\mu$ (Fig.~\ref{fig:4}{\bf a}-{\bf d}). The result is a family of scaling relations, where the footprint of small populations with high levels of mutation ({\it i.e.}, high noise, small limit cycle, Fig.~\ref{fig:4}{\bf c}) is approximately equivalent to that of large populations with low levels of mutation ({\it i.e.}, low noise, large limit cycle, Fig.~\ref{fig:4}{\bf d}). The {\it critical} scaling that defines the onset of the regime occurs when the stochastic footprint fills the mixed-strategy basin (Figs.~\ref{fig:4}{\bf a} \& {\bf e}), therefore facilitating the crossing of the separatrix.  Here, $O(\sqrt{\mu/N})$ fluctuations must be equivalent to the $O(\mu)$ deterministic repulsion in the direction normal to the AllC-TFT edge (Supplementary Information, Sec.~\ref{sec:boundary}), implying $N\sim1/\mu$, which agrees with both ensemble statistics and convex hull analysis ({\it cf.} Figs.~\ref{fig:3}{\bf a} \& \ref{fig:4}{\bf e}).

This behaviour can be recast as a size-dependent antagonistic relationship between mutation and intrinsic noise, where the former favours population heterogeneity (expelling towards the simplex boundaries) and the second homogeneity (attracting towards the centre of the simplex). This is particularly important in growing populations, since the balance between the two factors changes over time.

\subsection{Dynamical regimes in growing populations}\label{sec:growthandmutation}
\begin{figure}
\centering
\includegraphics[width=\columnwidth, trim=0 0 0 0, clip]{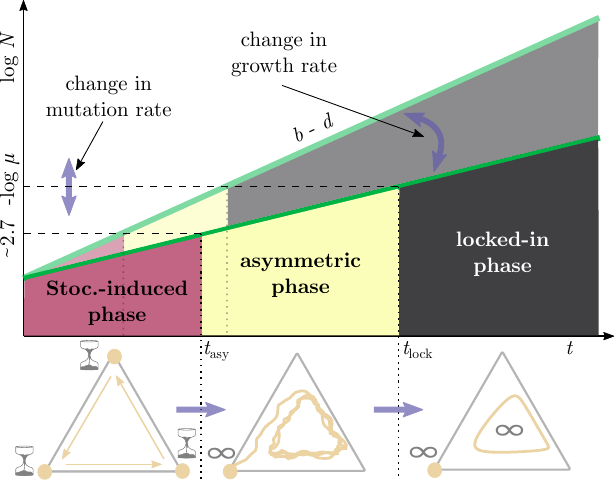}
\caption{
	\textbf{Growth and mutation control the relative duration of dynamical regimes.} By virtue of the size-dependent effective non-equilibrium phases (Fig.~\ref{fig:3}), a growing population exhibits three distinct dynamical regimes in sequence. Changing the overall rate of growth therefore re-scales the duration of the stochastically-induced and asymmetric phases. Changing the rate of mutation, by contrast, alters only the asymmetric phase. The duration of the asymmetric phases is important to overall fixation probabilities, since a longer duration implies a higher likelihood of crossing the separatrix, from which there is no return. 
}
\label{fig:new}
\end{figure}
As a population grows, it exhibits the three effective phases in sequence, with transitions occurring abruptly at critical population sizes (Fig.~\ref{fig:new}). Thus, despite the population size steadily increasing, dynamical behaviour manifests in three distinct regimes.

Seen through this lens, it is clear that changing the rate of growth changes the duration of the stochastically-induced and asymmetric phases by an overall factor (the locked-in phase is, in principle, open-ended). By contrast, the rate of mutation changes only the duration of the asymmetric phase (by virtue of the $\mu$-dependent transition to the locked-in phase).

The asymmetric phase is especially important for determining fixation probabilities. If the separatrix is crossed from the mixed-strategy basin during this phase, then the system remains in the basin of the AllD fixed point until the onset of the locked-in phase, where it remains for all until the population reaches $N_\mathrm{max}=10^{10}$.  Therefore, the longer the duration of the asymmetric phase, the greater the likelihood of crossing the separatrix. 

However, despite the appeal of this heuristic, we must also account for the fact that this likelihood is also conditioned on the state at which the system \emph{enters} the asymmetric regime. This is determined by the duration of preceding stochastically-induced regime, which is set by the rate of growth. It is also determined by the character of that regime, which is set by mutation. Despite the presence of two $\mu$-dependent mechanisms in the fixed-$N$ ensemble (see previous Section) it should be stressed that, in a {\it growing} system the duration of the stochastically-induced regime is sufficiently short in comparison to corner dwell times that, for all but pathologically slow growth rates or very large $\mu$, the separatrix is only ever likely to be crossed once. In turn, this puts greater emphasis on the role of the system's initial conditions.

To test our understanding of this complex interplay, we construct an approximation to the full outcome statistics. This is based on a decomposition in terms of conditional probabilities associated with each phase, and four `equivalence classes' of states at the start of the asymmetric phase.

\subsection{Equivalence classes}\label{sec:equiv}

\begin{figure*}
\centering
\includegraphics[width=\textwidth, trim=0 0 0 0, clip]{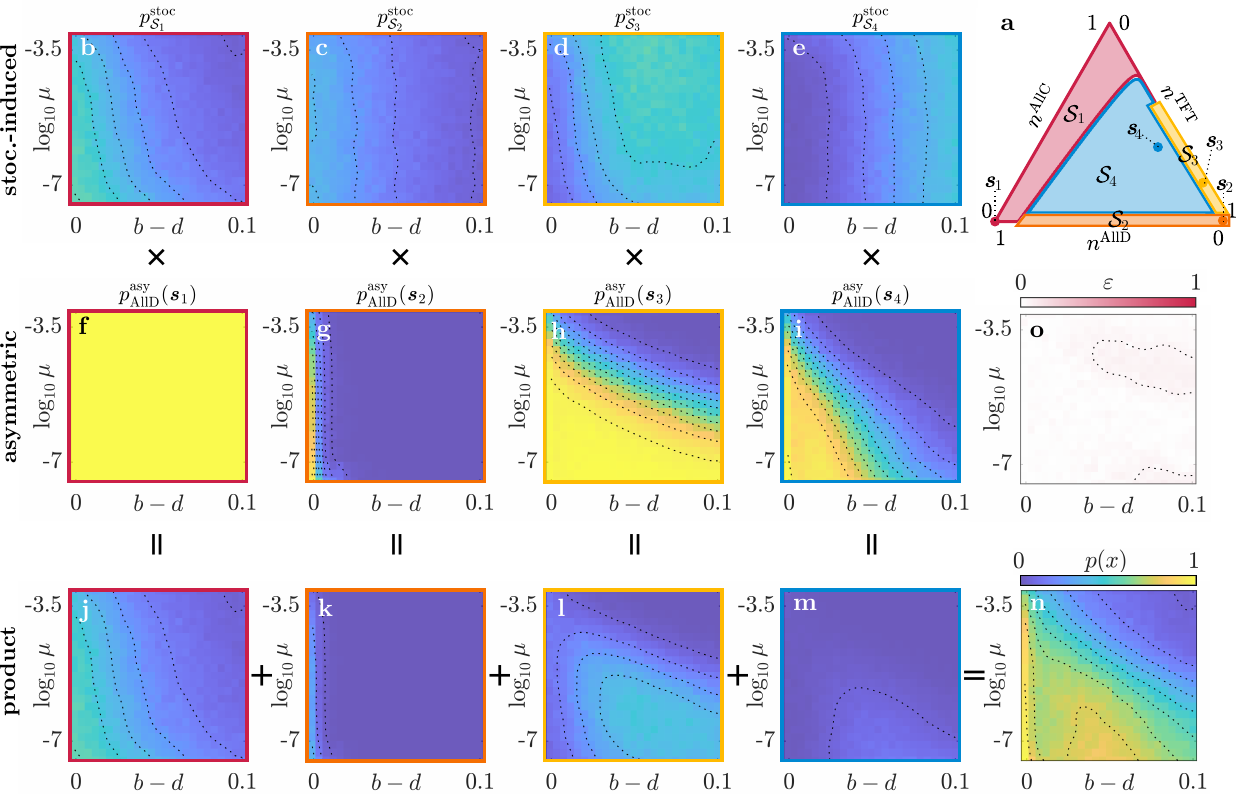}
\caption{
	\textbf{The dependence of fixation probabilities on rates of growth and mutation is reproduced by a decomposition based on effective phases and equivalence classes.}
	Fixation probabilities can be decomposed in terms of conditional probabilities that are based on equivalence classes of the asymmetric phase (panel \textbf{a}).  Within each class $\mathcal{S}_i$, $i=1,\ldots,3$, the outcome of the asymmetric phase does not rely on the specific demographic mix at the onset of the phase.  Moreover, the remaining states $\mathcal{S}_4$ contribute very little to the overall fixation probabilities (panels {\bf e}, {\bf i} \& {\bf m}, and main text).  This permits the approximation in Eq.~\eqref{eq:cond-2}, graphically represented by panels {\bf b}-{\bf n}.  The resulting reconstruction is in good agreement with the full stochastic simulations (panels \textbf{n} \& \textbf{o} and Fig.~\ref{fig:2}{\bf a}). The same decomposition is shown in Supplementary Information, Sec.~\ref{sec:additional-fix-b-m}, for additional initial conditions. The initial population size is $N_0=128$, whilst the times $t_\mathrm{asy}$ and $t_\mathrm{lock}$ are derived from the critical population sizes identified in Fig.~\ref{fig:3}.
}
\label{fig:5}
\end{figure*}

Consider the conditional probability $p^\text{asy}_\text{AllD}(\boldsymbol{s})=\Pr\left\{\boldsymbol{n}_{t_{\text{lock}}}\in\text{AllD-b}\,\vert\,\boldsymbol{n}_{t_{\text{asy}}} = \boldsymbol{s}\right\}$, where avoidance of extinction is now assumed implicitly. That is, the likelihood of being in the AllD basin at the onset of the locked-in phase, $t_{\text{lock}}$, given that the system was in a state $\boldsymbol{s}$ at the onset of the asymmetric phase, $t_{\text{asy}}$. Computing this probability via stochastic simulation demonstrates the existence of three approximate equivalence classes, $\mathcal{S}_i$, such that $p^\text{asy}_\text{AllD}(\boldsymbol{s}_i)$ is agnostic to the demographic mix $\boldsymbol{s}_i\in \mathcal{S}_i$, at the onset of the asymmetric phase (Fig.~\ref{fig:5}{\bf a} and Supplementary Information, Fig.~\ref{fig:areas-prob}). These are: those states in the AllD basin, $\mathcal{S}_1$; those along the AllD-TFT edge that stretch from the separatrix to the TFT corner, $\mathcal{S}_2$, and; those along the TFT-AllC edge (excluding the TFT corner) on the mixed-strategy side of the separatrix, $\mathcal{S}_3$.  The remaining states of the mixed-strategy basin are labelled $\mathcal{S}_4$.  Whilst these do not form an equivalence class, we assume (and later show) that they only minimally contribute to overall fixation probabilities.

The existence of equivalence classes prompts the following simplification (Supplementary Information, Sec.~\ref{sec:classes-dec}):
\begin{equation}
	\label{eq:cond-2}
		p_\mathrm{AllD}\approx \sum_{i=1}^4 p^\text{asy}_\text{AllD}(\boldsymbol{s}_i)\ p^\text{stoc}_{\mathcal{S}_i},
\end{equation}
where $p^\text{stoc}_{\mathcal{S}_i}=\Pr\left\{\boldsymbol{n}_{t_{\text{asy}}}\in \mathcal{S}_i\,\vert\,\boldsymbol{N}_0\right\}$ (again, with avoidance of extinction assumed implicitly), and states $\boldsymbol{s}_i$ can be chosen arbitrarily from $\mathcal{S}_i$.  This approximation allows us to verify our heuristic understanding of how the rates of growth and mutation impact fixation probabilities by controlling the duration and stochastic character of dynamical phases and therefore the likelihood of (stochastic) behaviours, such as crossing the separatrix, that are crucial in dictating long-term outcomes. It also dramatically reduces the computational time needed to calculate $p_\text{AllD}$ for a range of different initial conditions, since we only need to recalculate the four likelihoods that the stochastically-induced regime finishes in each of the equivalence classes, respectively.  The conditional probability $p_\text{AllD}^\text{asy}$, by contrast, does not depend on the specific initial conditions (but rather in which equivalence class the system is at time $t_\text{asy}$). 

\begin{figure*}[t]
\centering
\includegraphics[width=0.95\textwidth, trim=0 0 0 0, clip]{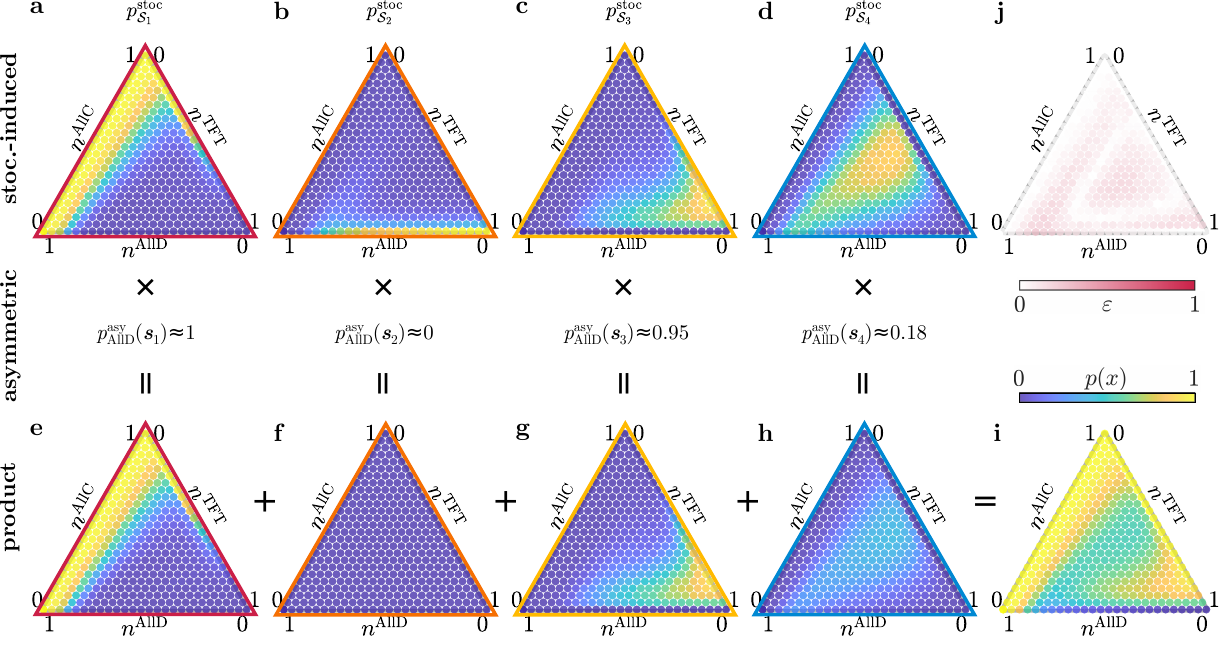}
\caption{
	\textbf{The sensitivity to initial conditions can be reproduced by a decomposition based on effective phases and equivalence classes.} For certain rates--- {\it i.e.}, those below the red line, Fig.~\ref{fig:2}{\bf a}--- fixation probabilities can rely critically on initial conditions.  That is, initial demographic mixes that are `close' (and not necessarily near the separatrix) can give rise to dramatically different long-term fixation probabilities.  This is particularly striking at low levels of mutation ($\log\mu=-6.5$) and comparatively high initial population size ($N_0=250$) ({\it cf.} Fig.~\ref{fig:2}{\bf f}).  The reason for this is twofold.  First, larger initial populations reduce the effect of the stochastically-induced phase, which is to expel populations to the simplex boundaries and the AllD and TFT corners (panels {\bf c} \& {\bf d}).  Second, the duration of the asymmetric phase is significant and cannot be neglected (middle row).  This implies that trajectories finishing the stochastically-induced phase in classes $\mathcal{S}_3$ or $\mathcal{S}_4$ can still cross the separatrix and fixate on the AllD fixed point (panels {\bf g} \& {\bf h}).  Whilst this provides a heuristic understanding, the contribution to $p_\text{AllD}$ from the $\mathcal{S}_4$ region violates one of the assumptions of (\ref{eq:cond-2}), and therefore the error increases with $N_0$ for the stated values of $\mu$ and $b-d$ (panel {\bf j} and Supplementary Information, Figs.~\ref{fig:dec-fix-N0-high-mu} \&~\ref{fig:dec-fix-N0-low-mu}).
	}  
\label{fig:6}
\end{figure*}

\subsubsection*{Growth and mutation} Fixing the mixed-strategy fixed point as the initial demographic mix, we can use Eq.~\eqref{eq:cond-2} to deconstruct the dependence of $p_\text{AllD}$ on the rates of growth and mutation (Fig.~\ref{fig:5}).

In the stochastically-induced regime, fluctuation gradients `drive' trajectories from the centre of the mixed-strategy basin towards the simplex edges and then the corners.  As a result, the growth-dependent (average) duration of the regime, $t^\text{asy} = \log{\left(10^{2.7}/N_0\right)}/(b-d)$, dictates in which class the trajectories are likely to start the asymmetric phase: rapid growth rates are required to confine trajectories that end in the mixed-strategy basin to the $\mathcal{S}_4$ region, whilst intermediate and slow growth rates suffice for the $\mathcal{S}_3$ and $\mathcal{S}_2$ regions, respectively (Figs.~\ref{fig:5}{\bf c}-{\bf e}). Although the average time spent in the two basins by such trajectories during the stochastically-induced regime is independent of $\mu$ (Fig.~\ref{fig:3}{\bf a}), the likelihood that the system is in ${\cal S}_1$ at time $t_\mathrm{asy}$ actually increases with decreasing $\mu$ (Fig.~\ref{fig:5}{\bf b}).  The reason is that decreasing $\mu$ changes the shape of the separatrix, therefore reducing the the size of the $\mathcal{S}_3$ region (Supplementary Information, Sec.~\ref{sec:stac-ind-detail}).

The asymmetric regime, by contrast, has a duration that is both growth- and $\mu$-dependent: $\tau_{\mu} = t_\mathrm{lock} - t_\mathrm{asy}=-\log{\left(\mu\,10^{2.7}\right)}/(b-d)$.  Here, the probability of crossing the separatrix hinges, principally, on the likelihood of avoiding the TFT corner and its associated large confinement times (Figs.~\ref{fig:5}{\bf g}-{\bf i}).  For example, $\tau^\text{asy}_\mu$ must be extremely long in order to permit crossings from the $\mathcal{S}_2$ region, since populations starting the asymmetric regime from this region encounter the TFT corner with almost certainty.  Crossings from the $\mathcal{S}_4$ region, however, occur at more modest $\tau^\text{asy}_\mu$--- achieved by either low growth and high $\mu$, or modest growth and low $\mu$--- reflecting the possibility that trajectories might avoid the TFT corner.  Those from the $\mathcal{S}_3$ region can happen at the smallest $\tau^\text{asy}_\mu$, since there is a high likelihood that trajectories will avoid the TFT corner (recall the anti-clockwise dynamics) and the ratio of the magnitude of stochastic effects to the magnitude of the deterministic flow is large in $\mathcal{S}_3$ (see Fig.~\ref{fig:3}\textbf{f}).

Combining these conditional probabilities using Eq.~\eqref{eq:cond-2} satisfactorily reproduces the overall statistics of demographic outcomes (\textit{cf.} Fig.~\ref{fig:5}{\bf n} and Fig.~\ref{fig:2}{\bf a}): the difference between our approximation and the full simulations have a mean value of \num{0.027}, when averaged over growth and mutation rates, and a maximum value of \num{0.099} (Fig.~\ref{fig:5}{\bf o}).  This also confirms our assertion that the trajectories that start the asymmetric phase from the $\mathcal{S}_4$ region do not impact long-term outcomes.  The reason is that this only happens with significant likelihood when growth rates are high, which simultaneously ensures that such trajectories never cross the separatrix (Figs.~\ref{fig:5}{\bf e}, {\bf i} \& {\bf m}). The same decomposition is shown for initial demographic mixes other than the mixed-strategy fixed point in Supplementary, Sec.~\ref{sec:additional-fix-b-m}.

\subsubsection*{Initial conditions} The approximation in Eq.~\eqref{eq:cond-2} further provides insight into the founder-like dependence of fixation probabilities on initial conditions (Fig.~\ref{fig:6} and Supplementary Information, Figs.~\ref{fig:dec-high-mu-low-pop}-\ref{fig:dec-low-mu-low-pop}). At high values of $\mu$, the asymmetric phase has negligible duration and behaviour is trivial (Figs.~\ref{fig:2}{\bf b} \& {\bf c}, Supplementary Information, Figs.~\ref{fig:dec-high-mu-low-pop} \& \ref{fig:dec-high-mu-high-pop}).  At low values of $\mu$, however, the asymmetric regime cannot be ignored and has a significant bearing on fixation probabilities.

In this case, if the initial population size is small (Fig.~\ref{fig:2}{\bf e}, Supplementary Information, Fig.~\ref{fig:dec-low-mu-low-pop}), then the stochastically-induced phase is sufficiently long as to expel trajectories to the AllD or TFT corners from the AllD or mixed-strategy basins, respectively.  Only the former trajectories impact fixation probabilities, however, since for all but the slowest growth rates, trajectories stuck in the TFT corner have a residence time longer than the duration of the asymmetric regime.

By contrast, if the initial population size is large (Fig.~\ref{fig:2}{\bf f} and Fig.~\ref{fig:6}), then the duration of the stochastically-induced phase is not long enough to expel trajectories to the boundaries and/or corners, resulting in a non-zero likelihood of starting the asymmetric regime from either the $\mathcal{S}_3$ or $\mathcal{S}_4$ regions.  For the former, there is a high probability of crossing to the AllD basin during the asymmetric regime, since the deterministic flows direct demographic trajectories towards the separatrix.  For the latter, this probability is much lower, since trajectories are more likely to be entrained to the limit cycle.  Nevertheless, the small contribution that results from the $\mathcal{S}_4$ region breaks one of the assumptions on which Eq.~\eqref{eq:cond-2} is based, which also explains why the error in our decomposition increases with $N_0$ for certain values of $\mu$ and $b-d$ (the mean value of the error in Fig.~\ref{fig:6}\textbf{j} is \num{0.079}, while the maximum value is \num{0.285}).  Of note, the values shown in Fig.~\ref{fig:6} are the worst case of those we have simulated (Supplementary Information, Sec.~\ref{sec:additional-no-mix}).

\section{Discussion}

Using a growing variant of an iconic model of evolutionary game theory, we have demonstrated the existence of non-trivial fluctuation-mediated effects in growing populations, whereby the rates of growth and mutation critically determine how fixation probabilities are conditioned on initial population size and demographic mix.  This has ramifications for the understanding of population bottlenecks and their long-term impact. The implication is that the population growth commonly associated with post-bottleneck recovery can, in fact, be as important as the effects of the bottleneck on population size and demographic mix. 

Our findings apply to populations that are well-mixed, and as such pertain to systems that combine short-range interactions with a mechanism for mixing that is fast on the timescales of the population dynamics, or those that otherwise have effectively long-range interactions, either explicitly or via the mutual interaction with public resource (although some of these assumptions have recently been brought into question \cite{Herrerias-Azcue2018a}).

In this context, we follow several pioneering works that have characterised various effects of growth in well-mixed populations \cite{Melbinger2010, Cremer2011, Huang2015, Constable2016, Ashcroft2017}. Although they do not explicitly examine the role of initial conditions, the behaviours appearing in two of these works \cite{Constable2016, Ashcroft2017} are related to the those reported here. As are several studies concerning growing systems of binary `spins' \cite{RGMTCR14,Klymko2017,Jack2019}.

The shared mechanism in all these cases is the presence of $O(1\sqrt{N})$ intrinsic fluctuations that decrease as the system grows. This has two main ramifications: first, ergodicity is broken, implying that a population fixates even in the presence of non-zero mutation, and; second, there is a decreasing scale by which fluctuations can couple dynamically to the geometry of underlying deterministic attractors of the system. It is the combination of these effects that gives rise to our headline behaviour: fixation probabilities that depend non-trivially on initial conditions (or otherwise stochastic events at early times), even as $t\to\infty$.

Such fluctuation-mediated effects are further characterised by a dependence on the rates of growth and mutation, which we explain by putting forward the notion of \emph{effective} non-equilibrium phase transitions, and showing that these delimit distinct demographic regimes in our model. The rate of growth controls the overall rate at which intrinsic fluctuations decrease, and therefore also the relative duration of such demographic regimes. Mutation, by contrast, has two related effects. First, it changes the nature of the underlying deterministic attractors (the size of the mixed-strategy limit cycle, in our case) and hence the structure to which decreasing fluctuations couple. Second, it also changes the state-dependence of the fluctuations. Whilst the former impacts the character of the initial stochastically-induced phase, it is the combination of both of these effects that sets the $\sim 1/\mu$ dependence of the critical transition between the intermediate asymmetric and the final locked-in phases.

The latter behaviour is an example of a potentially interesting and unexpected antagonistic relationship between two sources of stochasticity: mutation and the intrinsic effects of finite-sized populations. In our model and other studies, mutation promotes heterogeneity, whilst intrinsic fluctuations typically drive the system towards homogeneity \cite{RGMTCR14}. This is particularly important in the context of population growth, since the effects of mutation do not depend on the population size, whereas the stochastically-induced forcing due to intrinsic noise decreases with increasing population size. As a result, we speculate the other growing systems may also exhibit mutation-dependent critical transitions, where the effects of intrinsic fluctuations are balanced by those of mutation.

More generally, for growing well-mixed populations with all but the simplest of deterministic attractors---i.e., fixed points, limit cycles and stable manifolds etc.---the implication is that growth may be synonymous with effective non-equilibrium phase transitions.  Understanding whether and how these fit into the existing literature is an open question. Of potential interest is the ongoing challenge to classify non-equilibrium phase transitions by their universality classes \cite{Odor2004a}. Work has been undertaken to describe single absorbing state transitions (of the directed-percolation type) \cite{Hinrichsen2000a,Janssen1981, Grassberger1981} and also symmetric absorbing state transitions \cite{AlHammal2005}, but there appears to be very little literature on asymmetric absorbing state transitions. 

Theoretical considerations aside, we put forward that our ideas may be examined within the context of directed evolution \cite{romero2009exploring,tracewell2009directed}. Here, another, albeit direct, interplay between growth and mutation has already been demonstrated: mutations occurring at the genetic loci associated with growth-control promote so-called genetic instabilities \cite{coelho2019heterozygous}. Our results also appear relevant to state-of-the-art \emph{in silico} representations of directed evolution, where the role of intrinsic fluctuations during growth has so far been overlooked \cite{chang2021engineering}. A further setting that may prove relevant is that of quasispecies such as viruses. Viruses have both high rates of mutation and growth (\textit{cf.} Fig.~\ref{fig:2}{\bf a}, upper region), and appear to be largely unaffected by the many population bottlenecks involved in host-to-host transmission, as well as its intra-host (e.g., plaque-to-plaque) spreading \cite{domingo2012viral, domingo2021historical, weaver2021population}.

Nevertheless, bridging the gap between the abstract setting of the present work and the aforementioned applications will undoubtedly involve significant work. The extent to which this will be possible remains an open question, and may hinge on features that are not included in our model, such as spatial structure and/or other physical constraints \cite{lieberman2005evolutionary,allesina2011competitive,hindersin2015most,marrec2021toward}. So-called ``patch'' models of interacting locally well-mixed subpopulations is one avenue that may prove promising. Exploring how the ideas set out here translate across a wider class of systems is therefore an important avenue of future research, and we welcome further work in the area.

\section*{Author contributions}
EC and JP performed simulations and analysis, under the guidance of ST and RM.  EC and RM wrote the manuscript, with help from JP and ST. The project initially arose from discussions between JP, ST and RM.

\begin{acknowledgments}
The authors would like to thank V.~Guttal for helpful discussions at the project's outset.
RGM and EC acknowledge EMBL Australia for funding. We further acknowledge support from the Simons Foundation (Grant No.~287975 to ST), the Max Planck Society through a Max-Planck-Partner-Group at NCBS-TIFR (ST) and the Department of Atomic Energy, Government of India, under projects RTI4001 and RTI4006. This research includes computations using the computational cluster Katana supported by Research Technology Services at UNSW Sydney.
\end{acknowledgments}

\bibliographystyle{naturemag}

\appendix
\begin{widetext}

\newpage
\section*{Supplementary Information}

\subsection{Gillespie-It\^{o} simulations}
\label{sec:simulations}

The Gillespie algorithm \cite{gillespie1977exact,Gillespie2000a} allows the exact simulation of the stochastic dynamics of the growing IPD when $N$ is small, {\it i.e.}, when such dynamics cannot be approximated by simply integrating the SDEs (see Secs.~\ref{sec:sse} \&~\ref{sec:mul-noise}).
The downside of this algorithm is that its computational time scales linearly with $N$, becoming impractical as the population grows.
We therefore adopt a hybrid approach: when $N$ is smaller than a chosen threshold the system's dynamics are simulated with the Gillespie algorithm, and when the threshold is exceeded they are simulated by numerically integrating the SDEs (Euler-Maruyama).  Since the latter method results in values of $N^i$ along the real line, rounding is required if stochastic fluctuations trigger a switch back to the Gillespie algorithm.

Our choice for the algorithm switch threshold is $N=\mu^{-1}$.
This is motivated by the existence of the locked-in phase, which begins when $N \approx \mu^{-1}$ (see Fig.~\ref{fig:3}). During this phase, the risk that the SDE approximation would lead to an `accidental' crossing of the separatrix is extremely small.
A minimum threshold of $N=\num{10000}$ is used for simulations with large $\mu$.

All results were obtained using the High Performance Computing facility Katana \cite{Smith_Betbeder_Matibet_2010}.
The results in Fig.~\ref{fig:2}\textbf{a} (and Figs.~\ref{fig:cond-prop-centre}-\ref{fig:cond-prop-AllD}, panel \textbf{p}), were obtained with \num{10000} repetitions of the hybrid Gillespie-It\^{o} simulations for each combination of $b$ and $\mu$ (fixing $d=1$).
Each repetition was carried out until $N>10^{10}$.
The results in Fig.~\ref{fig:2}\textbf{b}-\textbf{g} were similarly obtained with \num{1000} repetitions for each initial condition.
The results in Fig.~\ref{fig:5}\textbf{b}-\textbf{i} (and Figs.~\ref{fig:cond-prop-centre}-\ref{fig:cond-prop-AllD}, panels \textbf{b}-\textbf{i}) were obtained with \num{1000} repetitions for each combination of $\mu$ and $b$.
\num{1000} repetitions were also used to obtain the results in Figs.~\ref{fig:6}\textbf{a}-\textbf{d} and Sec.~\ref{sec:additional-no-mix}.

The fixed size dynamics were also simulated using the hybrid Gillespie-It\^{o} approach.
The Gillespie algorithm was modified by setting $d=0$ and `killing' a randomly selected player at every birth.
The It\^{o} part of the algorithm involved replacing the noise correlation matrix (see details in Sec.~\ref{sec:lan-eq-fix-size}).
The results in Fig.~\ref{fig:3} were obtained by running the fixed size Gillespie-It\^{o} algorithm \num{1000} times for \num{20000} seconds for each combination of $\mu$ and $N$.

\subsection{The deterministic limit}
\label{sec:det-lim}

\begin{figure}[b!]
\includegraphics[width=0.89\textwidth, trim=0 0 0 2.7em]{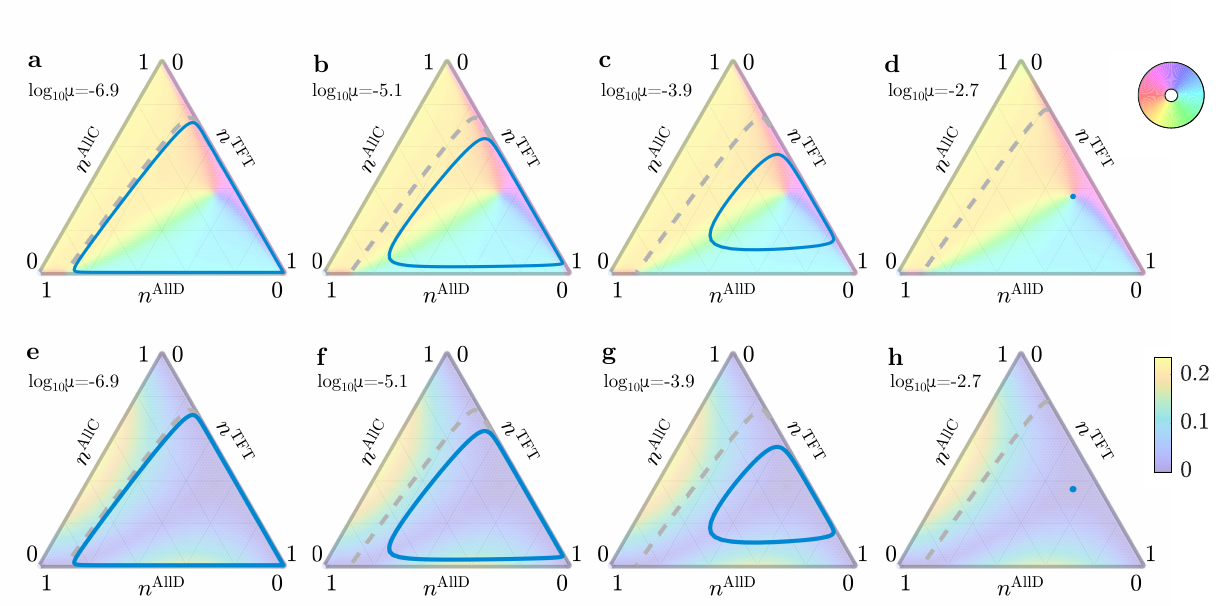}%
\caption{\textbf{Population dynamics in the deterministic limit of the growing IPD at different values of $\boldsymbol\mu$}.
The top row (panels \textbf{a}-\textbf{d}) show the direction of the determinist flow, while the bottom row (panels \textbf{e}-\textbf{h}) show the magnitude (speed).
For all panels, the growth rate is $b-d=0.05$.}
\label{fig:effect-mu}
\end{figure}

In the $N\to\infty$ limit, the behaviour of the growing Iterated Prisoner Dilemma described in the main text is equivalent to that of the following continuous, deterministic system:
\begin{equation}
\frac{d n^i}{d t} = \frac{b\mu}{\varphi}\sum_{j\neq i}(f_j n^j - f_i n^i) + \left (\frac{bf_i}{\varphi}-d\right )n^i,
\label{eq:effect-mu}
\end{equation}
where $n^i = N^i/N$.
These equations can be easily derived from Fig.~\ref{fig:setup}.

As shown in Fig.~\ref{fig:effect-mu} (see also~\cite{Imhof2005}), the mutation rate has a clear effect on the dynamics of the continuous system.
For $10^{-7.5}\,{\leq}\,\mu\,{\leq}\,10^{-2.5}$, the system exhibits two stable attractors: a fixed point characterised by a population of almost entirely AllD players and a stable mixed-strategy limit cycle.
The most notable effect of varying $\mu$ within this range is the resizing of the limit cycle: the smaller the mutation rate, the larger the limit cycle (panels \textbf{a}-\textbf{c} and \textbf{e}-\textbf{g}).
For large values of $\mu$ (approximately $10^{-3.1}\le\mu\le 10^{-2.5}$), the limit cycle collapses into a point, although the separatrix still exists (panels \textbf{d} and \textbf{h}).
Varying $\mu$ also slightly changes the shape of the separatrix, especially in the proximity of the AllC-TFT edge of the simplex.


\subsection{System-Size Expansion}
\label{sec:sse}

Following \cite{van_kampen,Gar03}, we derive a set of coupled SDEs that approximate the dynamics of the underlying protocol when $N\gg 1$. In terms of step-operators, the Master-equation has the form:
\begin{equation}
	\begin{split}
		\frac{dP(\vec{N};t)}{dt} &=\sum_i\left(\mathbb{E}^{-1}_{N^i} - 1\right) \frac{b f_i N^i}{\varphi} P(\vec{N};t)\\
		&\quad + \sum_i\left(\mathbb{E}^{+1}_{N^i} - 1\right) d\,N^i P(\vec{N};t)\\
		&\quad + \sum_i\sum_{j\neq i}\left(\mathbb{E}^{+1}_{N^i}\,\mathbb{E}^{-1}_{N^j} - 1\right)\frac{b \mu f_i N^i}{\varphi} P(\vec{N};t).
	\label{eq:ME}
	\end{split}
\end{equation}
Expanding the step-operators in the usual fashion, and retaining only the leading and next-to-leading order terms, gives 
\begin{equation}
	\mathbb{E}^r_{N^i} = 1 +
	 r \frac{\partial}{\partial N^i}
        + \frac{r^2}{2}\frac{\partial^2}{\partial (N^i)^2} + O\left(1\right),
\label{eq:expand}
\end{equation}
where $r=\pm 1$. Substituting Eq.~\eqref{eq:expand} into Eq.~\eqref{eq:ME} results in an equation of the Fokker-Planck type \cite{Gar03}:
\begin{equation}
\begin{split}
	\frac{\partial P(\vec{N};t)}{\partial t} &=
   	- \sum_i\frac{\partial}{\partial N^i}\Big[A_i(\vec{N})\,P(\vec{N};t)\Big]\\
   	&\quad + \frac{1}{2}\sum_i\sum_j\frac{\partial^2}{\partial N^i \partial N^j}\Big[B_{ij}\left(\vec{N}\right)\,P(\vec{N};t)\Big].
   	\end{split}
   	\label{eq:FP}
\end{equation}
After some manipulation, it can be shown that
\begin{equation}
	A_i = \frac{b\left(1-2\mu\right) f_i N^i}{\varphi} 
		+ \sum_{j\neq i} \frac{b\,\mu f_j N^j}{\varphi} - d\,N^i,
	\label{eq:alpha}
\end{equation}
and
\begin{equation}
	B_{ij} = 
	\left\{
	\begin{array}{cr}
		\frac{b\left(1+2\mu\right) f_i N^i}{\varphi} 
		+ \sum_{k\neq i} \frac{b\,\mu f_k N^k}{\varphi} + d\,N^i &\ \mathrm{if}\ i=j\\
		-\frac{b\mu}{\varphi}\left(f_i N^i + f_j N^j\right) &\ \mathrm{if}\ i\neq j 
	\end{array}
	\right. .
	\label{eq:beta}
\end{equation}
Equation \eqref{eq:FP} implies an SDE for the variables $N^i$:
\begin{equation}
	\frac{d N^i}{d t} = A_i + \xi_i .
	\label{eq:N_i}
\end{equation} 
The deterministic part of Eq.~\eqref{eq:N_i} is equivalent to Eq.~\eqref{eq:effect-mu}. The noise sources that appear in Eq.~\eqref{eq:N_i} have zero mean (\textit{i.e.}, $\langle\xi_i\rangle = 0$) and are correlated according to:
\begin{equation}
	\langle\xi_i(t)\xi_j(t^\prime)\rangle = B_{ij} \delta(t-t^\prime) .
\end{equation}
Summing over index $i$ in Eq.~\eqref{eq:N_i}, we have
\begin{equation}
	\frac{dN}{dt} = \left(b-d\right)N + \sqrt{\left(b+d\right)N}\xi,
	\label{eq:dN_dt}
\end{equation}
where $\xi$ is a single source of zero-mean Gaussian white noise.


\subsection{Multiplicative delta-correlated noise}
\label{sec:mul-noise}

Eqn.~\eqref{eq:N_i} can be re-written in terms of delta-correlated noise and multiplicative pre-factors.  To do this we must choose a matrix $\mathsf{b}$ that satisfies $\mathsf{B} = \mathsf{b}^T\cdot\mathsf{b}$.  For $\mathsf{b}$ to be square, this requires a Cholesky decompostion, and ensures no more independent noise sources than there are variables in the system.  However, a more natural approach is to decompose $\mathsf{B}$ according to the rules set out in Gillespie's CLE approach \cite{Gillespie2000a}.  This results is SDEs of the following form:
\begin{equation}
	\frac{d N^i}{d t} = A_i + \sum_\alpha b_{\alpha i}\eta_\alpha,
	\label{eq:N_i_mult}
\end{equation} 
where the index $\alpha$ runs from $1\ldots12$ and correspond to the reactions in Fig.~\ref{fig:setup}.  The twelve independent noise sources each have mean zero, and are delta-correlated--- \textit{i.e.}, $\langle \eta_\alpha\rangle = 0$ and $\langle \eta_\alpha(t)\eta_\beta(t^\prime)\rangle = \delta_{\alpha\beta}\delta(t-t^\prime)$. The matrix $\mathsf{b}^T$ is given by
\begin{equation}
	b_{\alpha i} = \left(
	\begin{array}{ccc}
		\sqrt{\frac{b f_1 N^1}{\varphi}} & 0 & 0 \\
		0 & \sqrt{\frac{b f_2 N^2}{\varphi}} & 0 \\
		0 & 0 & \sqrt{\frac{b f_3 N^3}{\varphi}} \\
		-\sqrt{d\,N^1} & 0 & 0 \\
		0 & -\sqrt{d\,N^2} & 0 \\
		0 & 0 & -\sqrt{d\,N^3} \\
		-\sqrt{\frac{b \mu f_1 N^1}{\varphi}} & \sqrt{\frac{b \mu f_1 N^1}{\varphi}} & 0 \\
		\sqrt{\frac{b \mu f_2 N^2}{\varphi}} & -\sqrt{\frac{b \mu f_2 N^2}{\varphi}} & 0 \\
		0 & -\sqrt{\frac{b \mu f_2 N^2}{\varphi}} & \sqrt{\frac{b \mu f_2 N^2}{\varphi}} \\
		0 & \sqrt{\frac{b \mu f_3 N^3}{\varphi}} & -\sqrt{\frac{b \mu f_3 N^3}{\varphi}} \\
		-\sqrt{\frac{b \mu f_1 N^1}{\varphi}} & 0 & \sqrt{\frac{b \mu f_1 N^1}{\varphi}} \\
		\sqrt{\frac{b \mu f_3 N^3}{\varphi}} & 0 & -\sqrt{\frac{b \mu f_3 N^3}{\varphi}} 
	\end{array}
	\right).
	\label{eq:b}
\end{equation}


\subsection{Projected dynamics}
\label{sec:proj}

We wish to project the dynamics onto the unit simplex--- \textit{i.e.}, in terms of variables $n^i = N^i / \sum_i N^i$.  For this, we require the multivariate form of It\^{o}'s lemma \cite{Gar03}.  For finite $N$, we have
\begin{equation}
	\frac{d n^i}{dt} = \sum_k A_k\,\partial_k n^i + \frac{1}{2}\sum_{k,j} B_{kj} \partial_k \partial_j n^i + \sum_j\sum_\alpha b_{j\alpha} \left(\partial_j n^i\right) \eta_\alpha,
\label{eq:Ito_1}
\end{equation}
%
where the shorthand $\partial_i = \partial/\partial N^i$ has been used.  First, we deal with the deterministic parts.  Using
\begin{equation}
	\frac{\partial n^i}{\partial N^j} = \left\{
	\begin{array}{cr}
		\frac{1 - n^i}{N} & \ \mathrm{if}\ i=j\\
		-\frac{n^i}{N} & \ \mathrm{if}\ i\neq j
	\end{array}
	\right. ,
	\label{eq:dn_i_dN_j}
\end{equation}
alongside (\ref{eq:alpha}), it can be shown that
\begin{equation}
	\sum_{k} A_k \partial_k n^i = \frac{b\mu}{\varphi}\left(\sum_{j\neq i} f_j n^j-2f_i n^i\right) + b\left(\frac{f_i}{\varphi} - 1\right) n^i.
\label{eq:dn_dt_term1}
\end{equation}
Similarly, using
\begin{equation}
	\frac{\partial^2 n^i}{\partial N^j\partial N^k} = \left\{
	\begin{array}{cr}
		\frac{2\left(n^i - 1\right)}{N^2} & \ \mathrm{if}\ i=j=k\\
		\frac{2n^i - 1}{N^2} & \ \mathrm{if}\ j = i \neq k\ \mathrm{or}\ k = i \neq j\\
		\frac{2n^i}{N^2} & \ \mathrm{if}\ j \neq i\ \mathrm{and}\ k \neq i
	\end{array}
	\right. ,
	\label{eq:d2n_i_dN_j_N_k}
\end{equation}
gives
\begin{equation}
	\frac{1}{2}\sum_{k,j} B_{kj} \partial_k \partial_j n^i = -b\left(\frac{f_i}{\varphi} - 1\right) \frac{n^i}{N}.
\label{eq:dn_dt_term2}
\end{equation}
Notice that, when summed over $i$, both \eqref{eq:dn_dt_term1} and \eqref{eq:dn_dt_term2} are zero by virtue of the fact that $\sum_i f_i n^i / \varphi = 1$.  Also, trivially, \eqref{eq:dn_dt_term2} goes to zero in the deterministic $N\to\infty$ limit.  For the stochastic part of \eqref{eq:Ito_1}, define a new matrix $b^\dagger_{\alpha i} = \sum_j b_{\alpha j}\left(\partial_j n^i\right)$ such that \eqref{eq:Ito_1} can be re-cast in terms of correlated noise sources--- \textit{i.e.},
\begin{equation}
	\frac{d n^i}{dt} = \frac{b\mu}{\varphi}\left(\sum_{j\neq i} f_j n^j-2f_i n^i\right)
	+ b\left(\frac{f_i}{\varphi} - 1\right) n^i \left(1 - \frac{1}{N}\right) + \zeta_i,
\label{eq:Ito_2}
\end{equation}
with $\langle \zeta_i\rangle = 0$ and $\langle \zeta_i(t)\zeta_j(t^\prime)\rangle = B^\dagger_{ij}\delta\left(t-t^\prime\right)$, where
%
\begin{equation}
	B^\dagger_{ij} = \sum_\alpha b^\dagger_{\alpha i}b^\dagger_{\alpha j} = \left\{
	\begin{array}{cr}
		\frac{b\,n^i}{N} \left[n^i + \frac{f_i}{\varphi}\left(1-2n^i\right)\right] + \frac{b\,\mu}{N}\left(1+\frac{f_i n^i}{\varphi}\right) + \frac{d\,n^i}{N}\left(1-n^i\right) & \ \mathrm{if}\ i=j\\
		-\frac{b\,n^i\,n^j}{N}\left(\frac{f_i}{\varphi} + \frac{f_j}{\varphi} - 1\right) - \frac{b\,\mu}{N\,\varphi} \left( f_i n^i + f_j n^j\right) - \frac{d\,n^i\,n^j}{N} & \ \mathrm{if}\ i\neq j
	\end{array}
	\right.,
	\label{eq:B_dagger}
\end{equation}
To understand the impact of these correlations, \eqref{eq:B_dagger} can be computationally decomposed into an eigenbasis for different values of $n^i$.  This always results in one zero-eigenvalue eigenvector pointing perpendicular to the simplex.  The remaining in-simplex eigenvalues reveal that populations towards the centre of the simplex experience large uncorrelated fluctuations whilst, closer to the simplex boundary, correlations suppress fluctuations in the direction normal to the boundary (Fig.~\ref{fig:3}\textbf{b}-\textbf{c}, orange crosses). Moreover, the magnitude of along-boundary fluctuations decrease as a corner is approached.


\subsection{Boundary effects}
\label{sec:boundary}

Equation \eqref{eq:Ito_2} can be evaluated at the simplex edges, where we are particularly interested in the both deterministic drift and fluctuations in the direction of the bulk, which is captured by the dynamics of the strategy who's concentration is zero along a given edge.

\begin{itemize}
	\item AllD-AllC edge: setting $n^3=0$ and $n^2=1-n^1$, gives $dn^3/dt=b\mu + \zeta_3$, where
	\begin{equation}
		\left\langle\zeta_3\zeta_j\right\rangle =
		\begin{pmatrix}
			-\frac{b\,\mu\,n^1 \left[n^1 (R-S)+S\right]}{N \left\{P(n^1-1)^2 + n^1\left[n^1 (R-S-T)+S+T\right]\right\}}\\
			-\frac{b\,\mu  (n^1-1) \left[(n^1-1) P-n^1T)\right]}{N \left\{P(n^1-1)^2 + n^1\left[n^1 (R-S-T)+S+T\right]\right\}}\\
			\frac{b\,\mu}{N}
		\end{pmatrix}.
	\end{equation}
	\item AllC-TFT edge: setting $n^2=0$ and $n^1=1-n^3$, gives $dn^2/dt=b\mu + \zeta_2$, where
	\begin{equation}
		\left\langle\zeta_2\zeta_j\right\rangle =
		\begin{pmatrix}
			\frac{b\,\mu\,m (n^3-1) R}{N(m R-c\,n^3)}\\
			\frac{b\,\mu}{N}\\
			\frac{b\,\mu\,n^3 (m R-c)}{N(c\,n^3-m R)}
		\end{pmatrix}.
	\end{equation}
	\item TFT-AllD edge: setting $n^1=0$ and $n^3=1-n^2$, gives $dn^1/dt=b\mu + \zeta_1$, where
	\begin{equation}
		\left\langle\zeta_1\zeta_j\right\rangle =
		\begin{pmatrix}
			\frac{b\,\mu }{N}\\
			-\frac{b\,\mu\,n^2\left[P (m+n^2-1)+T(n^2 -1)\right]}{N \left\{c (n^2-1)-(n^2)^2 \left[(m-2) P-m R+S+T\right]+ n^2 \left[2 (m-1) P-2 mR+S+T\right]+m R\right\}}\\
			\frac{b\,\mu  (n^2-1) \left\{-c+n^2 \left[(m-1) P-m R+S\right]+m R\right\}}{N
	   \left\{c (n^2-1)-(n^2)^2 \left[(m-2) P-m R+S+T\right]+n^2 \left[2 (m-1) P-2 m R+S+T\right]+m
	   R\right\}}
		\end{pmatrix}.
	\end{equation}
\end{itemize}

In all three of the above cases, the deterministic repulsion from the edge is
$O\left(\mu\right)$.  The fluctuations (positive and negative) in the same direction are $O(\sqrt{\mu/N})$.  The implication, in the context of the convex hull analysis of the main manuscript, is that, very close to the AllD-TFT edge, fluctuations can overcome the deterministic forces only if $N\sim 1/\mu$.


\subsection{Langevin equation for the fixed size model}
\label{sec:lan-eq-fix-size}

The dynamics of the system were were simulated using the hybrid Gillespie-It\^{o} approach (see Sec.~\ref{sec:simulations}).
A fixed size version of the system was also used in our analysis (see Figs.~\ref{fig:3} \&~\ref{fig:4}): The Gillespie algorithm was modified by setting $d=0$ and `killing' a randomly selected player at every birth, while
the It\^{o} part of the algorithm involved replacing the matrix $\mathsf{b}^T$ in Eq.~\eqref{eq:b} with
\begin{equation}
	b_{\alpha i} = \left(
	\begin{array}{ccc}
		(1-n^1)\sqrt{\frac{b f_1 N^1}{\varphi}} & -n^1\sqrt{\frac{b f_2 N^2}{\varphi}} & -n^1\sqrt{\frac{b f_3 N^3}{\varphi}} \\
		-n^2\sqrt{\frac{b f_1 N^1}{\varphi}} & (1-n^2)\sqrt{\frac{b f_2 N^2}{\varphi}} & -n^2\sqrt{\frac{b f_3 N^3}{\varphi}} \\
		-n^3\sqrt{\frac{b f_1 N^1}{\varphi}} & -n^3\sqrt{\frac{b f_2 N^2}{\varphi}} & (1-n^3)\sqrt{\frac{b f_3 N^3}{\varphi}} \\
		0 & 0 & 0 \\
		0 & 0 & 0 \\
		0 & 0 & 0 \\
		-\sqrt{\frac{b \mu f_1 N^1}{\varphi}} & \sqrt{\frac{b \mu f_1 N^1}{\varphi}} & 0 \\
		\sqrt{\frac{b \mu f_2 N^2}{\varphi}} & -\sqrt{\frac{b \mu f_2 N^2}{\varphi}} & 0 \\
		0 & -\sqrt{\frac{b \mu f_2 N^2}{\varphi}} & \sqrt{\frac{b \mu f_2 N^2}{\varphi}} \\
		0 & \sqrt{\frac{b \mu f_3 N^3}{\varphi}} & -\sqrt{\frac{b \mu f_3 N^3}{\varphi}} \\
		-\sqrt{\frac{b \mu f_1 N^1}{\varphi}} & 0 & \sqrt{\frac{b \mu f_1 N^1}{\varphi}} \\
		\sqrt{\frac{b \mu f_3 N^3}{\varphi}} & 0 & -\sqrt{\frac{b \mu f_3 N^3}{\varphi}} 
	\end{array}
	\right).
\end{equation}


\subsection{Escape time from corners in the stochastically-induced regime}
\label{sec:escape-time}

\begin{figure}[h!]
\centering
\includegraphics[width=\textwidth, trim=0 0 0 0, clip]{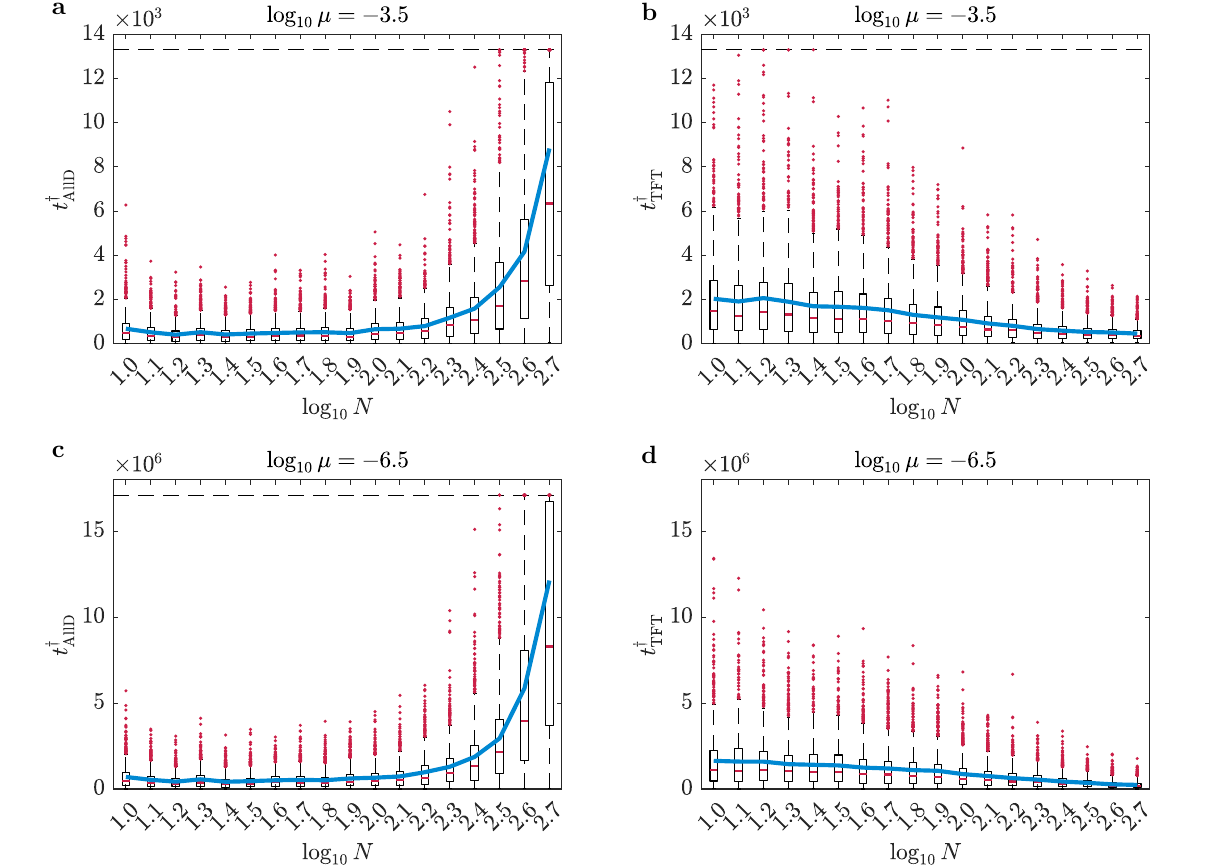}%
\caption{
\textbf{Statistics of separatrix crossing time in the stochastically induced regime.} The figure shows the statistics of the time taken by the system to cross the separatrix for the first time using the fixed-size model with $\boldsymbol{n}_0 = \boldsymbol{s}_1$, i.e. the AllD corner (panels \textbf{a} for $\log_{10}\mu=-3.5$ and \textbf{c} for $\log_{10}\mu=-6.5$) and with with $\boldsymbol{n}_0 = \boldsymbol{s}_2$, i.e. the TFT corner (panels \textbf{b} for $\log_{10}\mu=-3.5$ and \textbf{d} for $\log_{10}\mu=-6.5$). The statistics, for each of the four cases, are obtained from \num{10000} repetitions. On each box, the central red mark indicates the median, and the bottom and top edges indicate the 25th and 75th percentiles, respectively, the whiskers extend to the most extreme data points not considered outliers, and the red dots are the outliers (a data value is considered an outlier if it is greater than $Q_3 + 1.5 (Q_3 – Q_1)$ or less than $Q_1 – 1.5 (Q_3 – Q_1)$, where $Q_1$ and $Q_3$ are the 25th and 75th percentiles, respectively). For an easier visualisation of the statistics, all data values exceeding a maximum threshold are collapsed into the horizontal dashed lines on the top part of each plot. 
The blue lines represent the average crossing time over all repetitions.
}
\label{fig:ET}
\end{figure}
Here we show the statistics of the time $t^{\dagger}$ that is needed by the fixed-size system to cross the separatrix for the first time (Fig.~\ref{fig:ET}). We compare the cases of the systems starting from the AllD corner (panels \textbf{a} and \textbf{c}) and the TFT corner (panels \textbf{b} and \textbf{d}).
In both cases, $t^{\dagger}$ is proportional to $\mu$: in the order of thousands of seconds when $\log_{10}\mu=-3.5$ (panels \textbf{a-b}) and in the order of millions of seconds when $\log_{10}\mu=-6.5$ (panels \textbf{c-d}).
However, the influence of the population size $N_0$ on $t^{\dagger}$ is different in the two cases.
On the one hand, the crossing time starting from the AllD corner increases very rapidly when $N_0$ approaches the onset of the asymmetric regime at around $N_0=10^{2.7}$ (panels \textbf{a} and \textbf{c}), indeed showing that crossing the separatrix from the AllD basin to the mixed-strategy basin becomes extremely unlikely in the asymmetric regime.
On the other hand, the crossing time starting from the TFT corner decreases with $N_0$.


\subsection{Equivalence classes decomposition}
\label{sec:classes-dec}

\begin{figure}[t!]
\includegraphics[width=\textwidth, trim=2cm 0 2cm 0, clip]{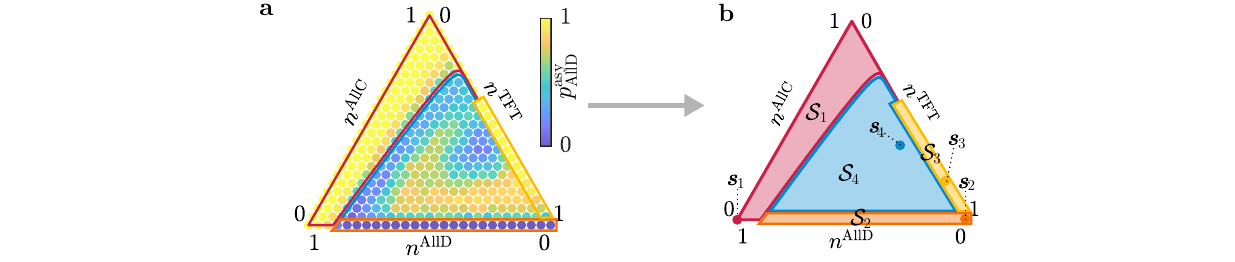}%
\caption{
\textbf{Equivalence classes of the asymmetric regime}. $p_\mathrm{AllD}^\mathrm{asy}=\mathrm{Pr} \{\boldsymbol{n}_{t_\mathrm{lock}} \in \textrm{AllD-b} \ | \ \boldsymbol{n}_{t_\mathrm{asy}} = \boldsymbol{s}\}$ was numerically estimated via Gillespi-It\^{o} simulations for a subset of all points $\boldsymbol{s}$ (panel \textbf{a}).
The results motivates the decomposition of the simplex into four equivalence classes, ${\cal S}_i$, as well as the choice of their representative point $\boldsymbol{s}_i$ (panel \textbf{b}).
The parameters used for this illustration are $b-d=0.05$ and $\log_{10}\mu=-6.5$.
}
\label{fig:areas-prob}
\end{figure}

Consider a trajectory of the growing IPD that starts from an arbitrary state in the stochastically-induced phase (i.e., $N_0<10^{2.7}$) at time $t=0$, enters the asymmetric phase at time $t=t_\textrm{asy}$ and the locked-in phase at time $t=t_\textrm{lock}$. Due to the nature of the locked-in phase, the evolutionary outcome at $t\,{\to}\,\infty$ is already known at time $t_\textrm{lock}$, and thus Eq.~\eqref{eq:palld} can be simplified:
\begin{equation}
p_\textrm{AllD}= \mathrm{Pr} \{\boldsymbol{n}_{t_\mathrm{lock}} \in \textrm{AllD-b}\,\vert\,\boldsymbol{N}_0\},
\end{equation}
where avoidance of extinction is implicitly assumed and $\textrm{AllD-b}$ indicates the $\textrm{AllD}$ basin of the simplex.

Moreover, since the growing IPD is a Markovian process, $p_\textrm{AllD}$ can be decomposed as follows:
\begin{equation}
\label{eq:cond-1}
p_\mathrm{AllD} = \sum_{\boldsymbol{s}\in U} \mathrm{Pr} \{\boldsymbol{n}_{t_\mathrm{lock}} \in \textrm{AllD-b} \ | \ \boldsymbol{n}_{t_\mathrm{asy}} = \boldsymbol{s}\}\ \mathrm{Pr} \{\boldsymbol{n}_{t_\mathrm{asy}} = \boldsymbol{s} \ | \ \boldsymbol{N}_0 \},
\end{equation}
where $U$ is the set of all possible states of the system at time $t_\textrm{asy}$.

Eq.~\eqref{eq:cond-1} is impractical, since it requires the numerical estimation of the probabilities $\mathrm{Pr} \{\boldsymbol{n}_{t_\mathrm{lock}} \in \textrm{AllD-b} \ | \ \boldsymbol{n}_{t_\mathrm{asy}} = \boldsymbol{s}\}$ and $\mathrm{Pr} \{\boldsymbol{n}_{t_\mathrm{asy}} = \boldsymbol{s} \ | \ \boldsymbol{N}_0 \}$ for all $\boldsymbol{s}$ in the very large set, $U$.
However, the approximation given in Eq.~\eqref{eq:cond-2} can be made by inspecting $\mathrm{Pr} \{\boldsymbol{n}_{t_\mathrm{lock}} \in \textrm{AllD-b} \ | \ \boldsymbol{n}_{t_\mathrm{asy}} = \boldsymbol{s}\}$ in Fig.~\ref{fig:areas-prob}\textbf{a}, which was numerically estimated for a subset of all points $\boldsymbol{s}$ (for illustration purposes we only show the case of $b-d=0.05$ and $\log_{10}\mu=-6.5$).

A first area of the simplex, ${\cal S}_1$ (outlined in red), corresponding to the $\textrm{AllD}$ basin, can be immediately identified as it is characterised by a homogenous probability of $\textrm{AllD}$ outcomes of approximately 1.
If the system is anywhere within ${\cal S}_1$ at time $t_\textrm{asy}$, then it is expected never to move to the mixed-strategy basin, since the crossing of the separatrix in this direction is extremely unlikely.
Any point $\boldsymbol{s}_1$ would be good candidate for representing the entire area ${\cal S}_1$, however, we chose the $\textrm{AllD}$ corner (red dot in Fig.~\ref{fig:areas-prob}\textbf{b}) as the system is in the proximity of such point for the vast majority of the time spent in ${\cal S}_1$.

A second area can similarly be identified: ${\cal S}_2$ (orange), corresponding to the mixed-strategy part of the $\textrm{AllD-TFT}$ edge (including the $\textrm{TFT}$ corner) and characterised by a homogenous probability of $\textrm{AllD}$ outcomes of approximately 0.
If the system is in this area at time $t_\textrm{asy}$, then it is very unlikely to cross the separatrix because of the long time (inversely proportional to $\mu$) spent in the $\textrm{TFT}$ corner while growing (i.e., as the fluctuations become smaller and smaller).
Again, any point $\boldsymbol{s}_2$ can represent the area ${\cal S}_2$, but we chose the $\textrm{TFT}$ corner (orange dot in Fig.~\ref{fig:areas-prob}\textbf{b}).

\begin{figure}[t!]
\includegraphics[width=\textwidth, trim=0 0 0 0, clip]{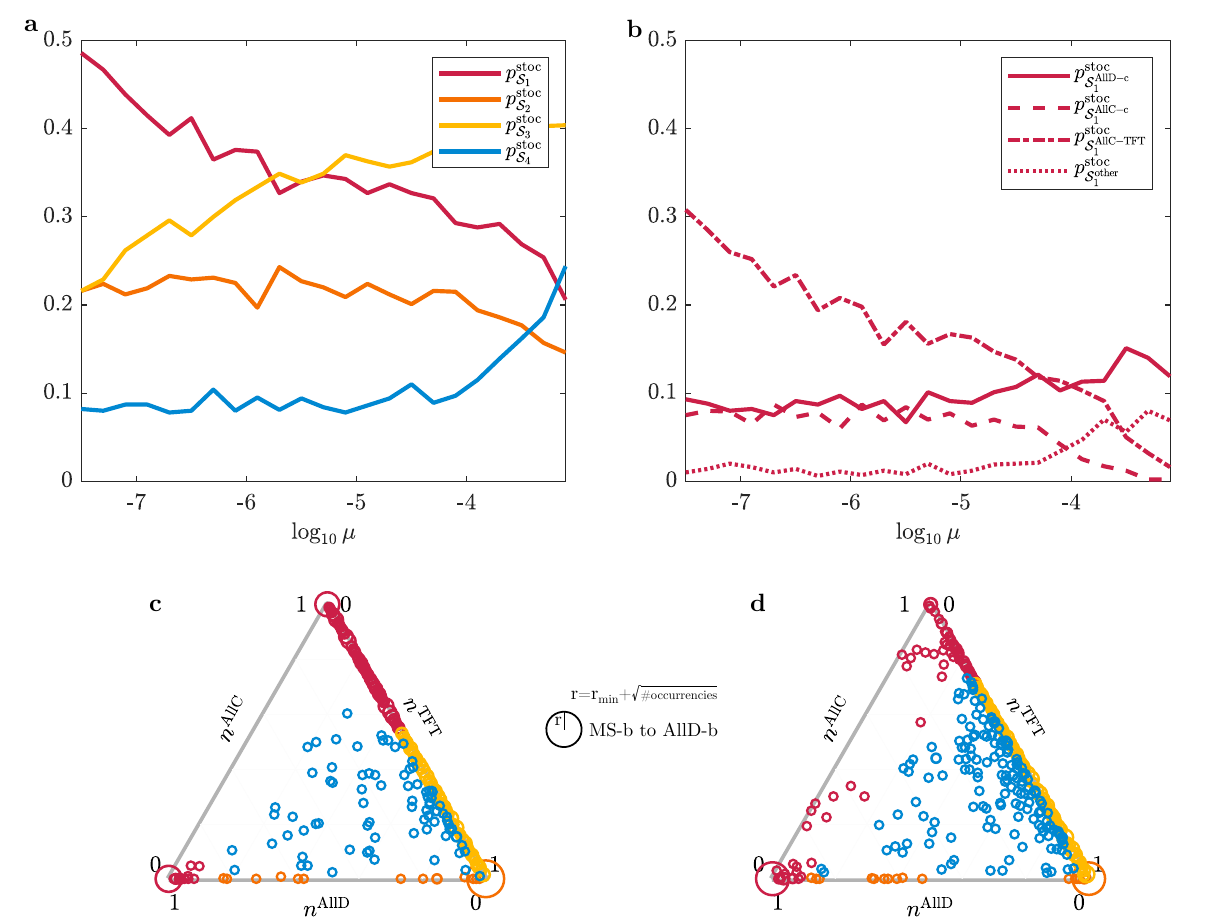}%
\caption{
\textbf{Statistics of the stochastically induced regime $\boldsymbol{p^\mathrm{stoc}_{{\cal S}_i}}$ in detail.}
Panel \textbf{a} shows the probabilities $p_{{\cal S}_i}^\mathrm{stoc}$ over $\mu$ with $b-d=0.026$, $\boldsymbol{n}_0=\textrm{MS-fp}$ and $N_0=128$.
Panel \textbf{b} decomposes $p_{{\cal S}_1}^\mathrm{stoc}$ into cases of the AllD corner, the AllC corner, the AllD basin side of the AllC-TFT edge (excluding the AllC corner) and the rest of the AllD basin.
Panels \textbf{c} and \textbf{d} show the location of $\boldsymbol{n}_{t_\mathrm{asy}}$ within the simplex for the same \num{1000} stochastic trajectories used for estimating the probabilities in \textbf{a} and \textbf{b}, for the cases of $\log_{10}\mu=-6.5$ and $\log_{10}\mu=-3.5$ respectively.
The size of the circles is proportional to the occurrences of $\boldsymbol{n}_{t_\mathrm{asy}}$ in the location and the colours reflect the different equivalence classes ${\cal S}_i$ (see panel \textbf{a}).
}
\label{fig:stoc-ind-det}
\end{figure}

A third area, ${\cal S}_3$ (yellow), corresponding to the mixed-strategy part of the $\textrm{AllC-TFT}$ edge with exclusion of the $\textrm{TFT}$ corner, also stands out: here we see the probability of $\textrm{AllD}$ outcomes quickly increases with the fraction of $\textrm{AllC}$ players.
Due to the anti-clockwise dynamics, in ${\cal S}_3$ the system directed towards the point where the separatrix meets the $\textrm{AllC-TFT}$ edge, which is where a move from the mixed-strategy basin to the $\textrm{AllD}$ basin is most likely (see Fig.~\ref{fig:3}\textbf{f}).
Since the probability of $\textrm{AllD}$ outcomes is less homogeneous, an approximation must be made for $\boldsymbol{s}_3$.
The dynamics along the $\textrm{AllC-TFT}$ edges are slow, and they get even slower close to the corners.
We observe that a good proxy for the average position of the system over time during a climb of the $\textrm{AllC-TFT}$ edge from the TFT corner to the separatrix is around $n^\textrm{AllC}=0.25$ and $n^\textrm{TFT}=0.75$ (yellow dot in Fig.~\ref{fig:areas-prob}\textbf{b}).

Within the remainder of the simplex, ${\cal S}_4$ (blue), $\mathrm{Pr} \{\boldsymbol{n}_{t_\mathrm{lock}} \in \textrm{AllD-b} \ | \ \boldsymbol{n}_{t_\mathrm{asy}} = \boldsymbol{s}\}$ can be very heterogeneous, however, as discussed in the main text, this contributes very little to the outcome statistics for small values of $N_0$.
We chose $\boldsymbol{s}_4=\textrm{MS-fp}$ to represented this area (blue dot in Fig.~\ref{fig:areas-prob}\textbf{b}).

Fig.~\ref{fig:areas-prob}\textbf{a} was obtained with $b-d=0.05$ and $\log_{10}\mu=-6.5$.
Different values of these parameters produce different values of $\mathrm{Pr} \{\boldsymbol{n}_{t_\mathrm{lock}} \in \textrm{AllD-b} \ | \ \boldsymbol{n}_{t_\mathrm{asy}} = \boldsymbol{s}\}$, however, the four areas can always be identified.
This decomposition is validated by the successful reconstruction (see Figs.~\ref{fig:5} \&~\ref{fig:6} as well as Figs~\ref{fig:cond-prop-centre}-\ref{fig:dec-fix-N0-low-mu}) of the all statistics of the evolutionary outcomes obtained via full simulations (Fig.~\ref{fig:2}).
The conditional probability decomposition is always very accurate except for the case of low mutation rate (e.g., $\log_{10}\mu<-6$), growth rates around $b-d \approx 0.05$ and large initial populations (e.g., $N_0>200$), for which it begins to produce less accurate results (Fig.~\ref{fig:6} panel \textbf{o}).


\subsection{Statistics of the stochastically-induced regime in detail}
\label{sec:stac-ind-detail}
In this section we describe the effect of the mutation rate on the statistics of the stochastically-induced regime outcomes more in detail.
Fig.~\ref{fig:stoc-ind-det}\textbf{a} shows the same probabilities $p_{{\cal S}_i}^\mathrm{stoc}$ shown in Figs.~\ref{fig:5}\textbf{b}-\textbf{e}, but for a single value of $b-d=0.026$ (we remind that $\boldsymbol{n}_0=\textrm{MS-fp}$ and $N_0=128$).
We can see that all probabilities depend on $\mu$ for high mutation rates (i.e., $\log_{10}\mu>-4$), but only $p_{{\cal S}_1}^\mathrm{stoc}$ and $p_{{\cal S}_3}^\mathrm{stoc}$ depend on $\mu$ for lower mutation rates.

For a more detailed analysis, we decompose $p_{{\cal S}_1}^\mathrm{stoc}$ into $p_{{\cal S}_1^\textrm{AllD-c}}^\mathrm{stoc}$, the probability of $\boldsymbol{n}_{t_\mathrm{asy}}$ being the AllD corner, $p_{{\cal S}_1^\textrm{AllC-c}}^\mathrm{stoc}$, the probability of $\boldsymbol{n}_{t_\mathrm{asy}}$ being the AllC corner, $p_{{\cal S}_1^\textrm{AllC-TFT}}^\mathrm{stoc}$, the probability of $\boldsymbol{n}_{t_\mathrm{asy}}$ being in the AllD basin side of the AllC-TFT edge excluding the AllC corner and $p_{{\cal S}_1^\textrm{other}}^\mathrm{stoc}$, the probability of $\boldsymbol{n}_{t_\mathrm{asy}}$ being anywhere else within the AllD basin.
We can now see that $p_{{\cal S}_1^\textrm{AllC-TFT}}^\mathrm{stoc}$ decreases with $\mu$, in contrast with $p_{{\cal S}_3}^\mathrm{stoc}$, which instead increases with $\mu$.
This is explained by the shape of the separatrix changing with $\mu$: the point where the separatrix meets the $\textrm{AllC-TFT}$ edge moves towards the TFT corner as $\mu$ decreases (see panels \textbf{c} \& \textbf{d}).

At the same time, we also see that the probability of the system being in one of the two corners at time $t_\textrm{asy}$, $p_{{\cal S}_1^\textrm{AllD-c}}$ and $p_{{\cal S}_1^\textrm{AllC-c}}$, are independent on the mutation rate for $\log_{10}\mu<-5$.
However, for higher mutation rates it becomes more and more likely for the system to be in the AllD corner rather than in the AllC corner (see also panels \textbf{c} \& \textbf{d}).
Finally, for mutation rates $\log_{10}\mu>-5$ the probability of the system being away from the boundaries of the simplex $p_{{\cal S}_1^\textrm{other}}^\mathrm{stoc}$ becomes higher (the same of course can be observed for $p_{{\cal S}_3}^\mathrm{stoc}$).


\subsection{Outcome statistics over growth/mutation rates: more initial mixes}
\label{sec:additional-fix-b-m}

Here we show that our decomposition of the outcome statistics based on equivalence classes holds not only for $\boldsymbol{n}_0=\textrm{MS-fp}$ but in general. We consider other three cases that are very different: $\boldsymbol{n}_0 = \textrm{centre}$, i.e., $n^\textrm{AllC} = n^\textrm{AllD} = n^\textrm{TFT} = 1/3$ (Fig.~\ref{fig:cond-prop-centre}); $\boldsymbol{n}_0=\textrm{TFT-c}$, i,e., $n^\textrm{AllC} = n^\textrm{AllD} = 0$, $n^\textrm{TFT} = 1$ (Fig.~\ref{fig:cond-prop-TFT}); and $\boldsymbol{n}_0=\textrm{AllD-c}$, i,e., $n^\textrm{AllC} = n^\textrm{TFT} = 0$, $n^\textrm{AllD} = 1$ (Fig.~\ref{fig:cond-prop-AllD}).
In all figures, comparing the results obtained via decomposition (panels \textbf{n}) against those obtained via full simulation (panels \textbf{p}) yields a low error (panels \textbf{o}), demonstrates the validity of our method.
The average error over all the considered values of growth and mutation is approximately \num{0.029} for the $\boldsymbol{n}_0 = \textrm{centre}$, \num{0.003} for $\boldsymbol{n}_0=\textrm{TFT-c}$, and \num{0.001} for $\boldsymbol{n}_0=\textrm{AllD-c}$.

\newpage
\begin{figure}[h!]
\includegraphics[width=0.9\textwidth, trim=0 0 0 0, clip]{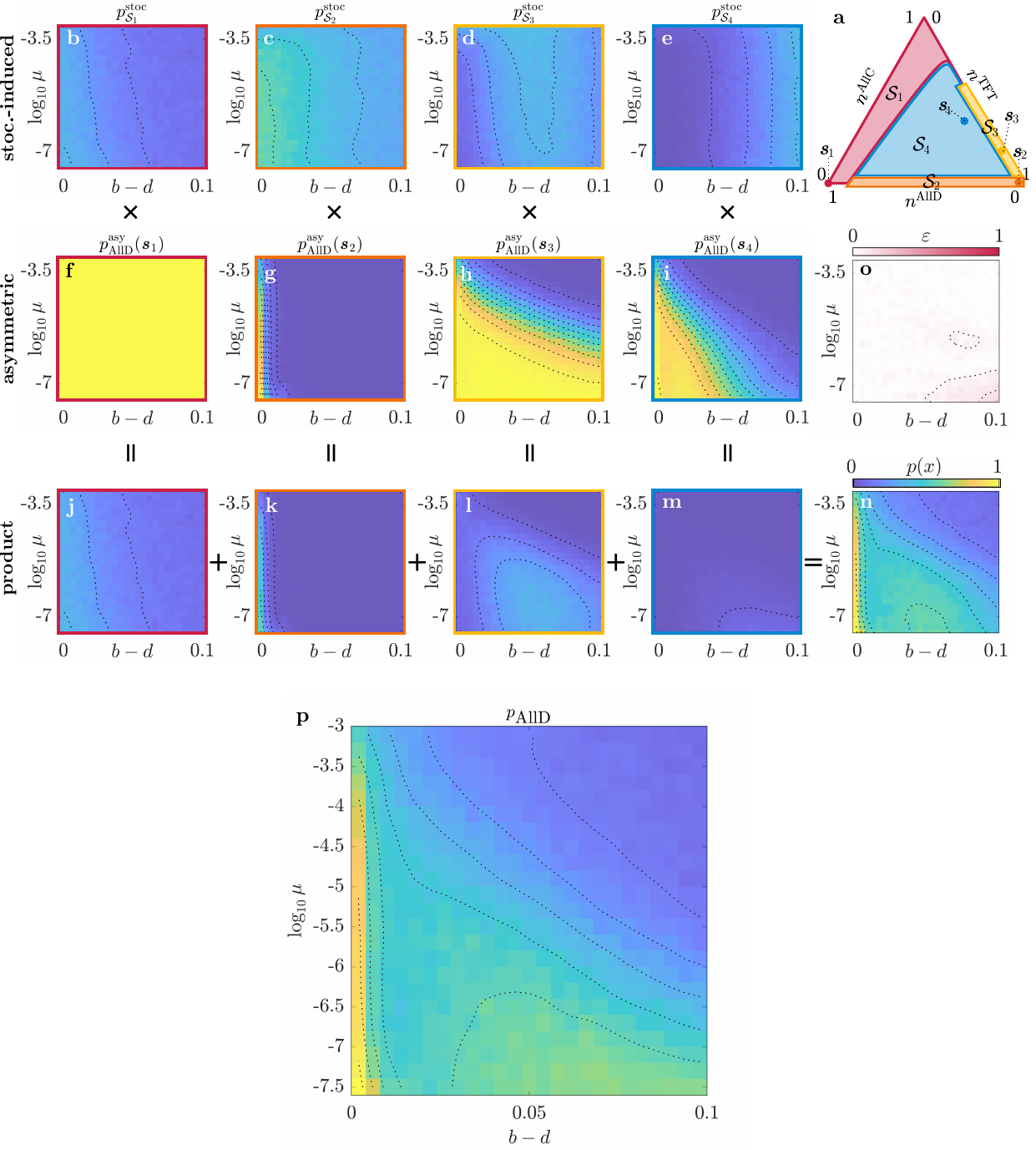}%
\caption{
\textbf{Decomposition of outcome statistics based on equivalence classes ($\boldsymbol{n_0=\mathrm{centre}}$).}
The statistics of evolutionary outcomes can be decomposed in terms of conditional probabilities that are based on four equivalence classes of states (panels \textbf{a}-\textbf{i}).
The landscape of $p_\mathrm{AllD}(\mathrm{centre})$ for different rates of growth ($b-d$) and mutation ($\mu$) can be reconstructed with good agreement (panels \textbf{j}-\textbf{n} and panel \textbf{o}).
The initial population size is $N_0 = 128$. $t_\mathrm{asy}$ and $t_\mathrm{lock}$ are derived from the critical population sizes identified in Fig.~\ref{fig:3}.
}
\label{fig:cond-prop-centre}
\end{figure}

\newpage
\begin{figure}[h!]
\includegraphics[width=0.9\textwidth, trim=0 0 0 0, clip]{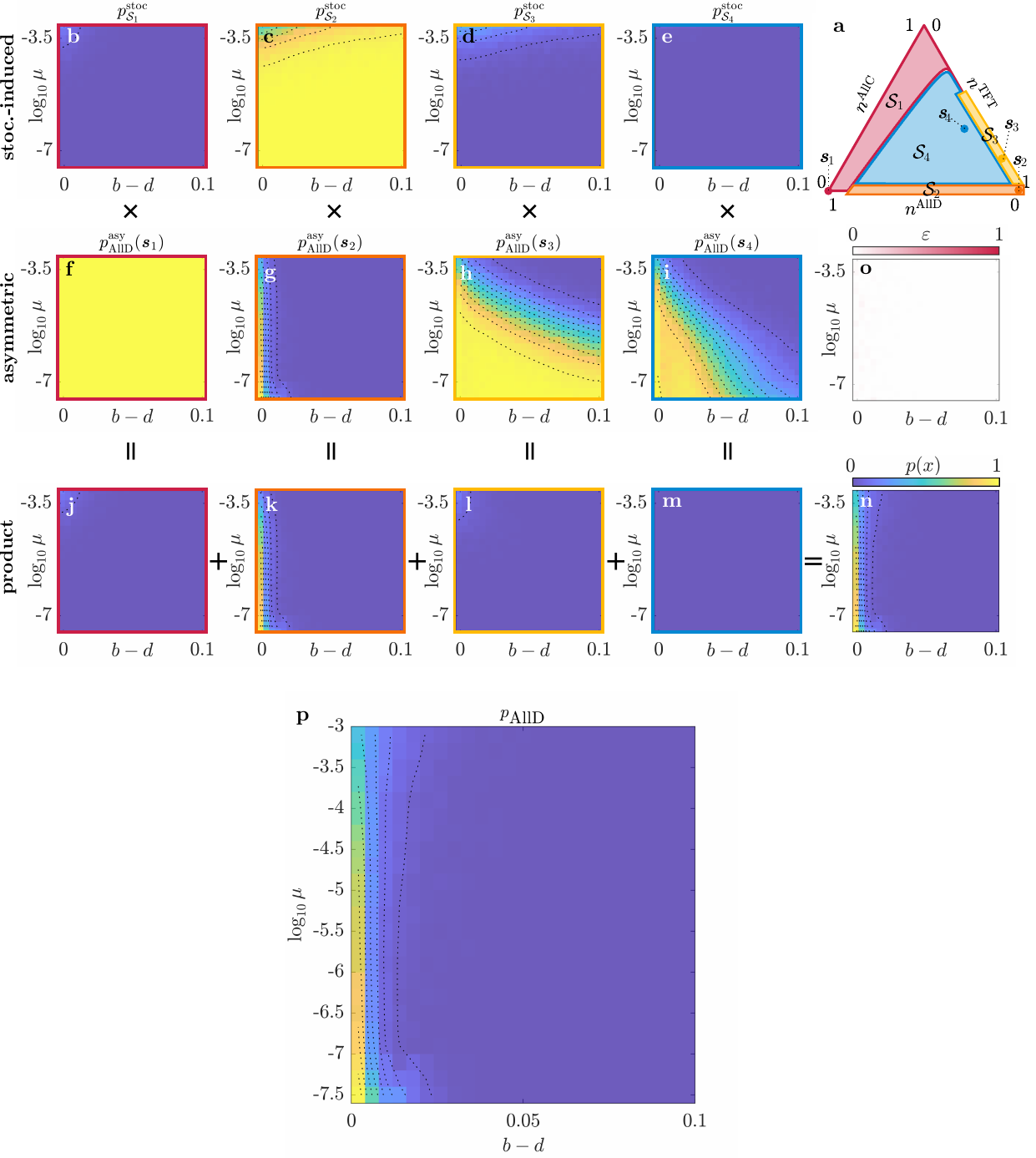}%
\caption{\textbf{Decomposition of outcome statistics based on equivalence classes ($\boldsymbol{n_0=\textrm{TFT-c}}$).}
The statistics of evolutionary outcomes can be decomposed in terms of conditional probabilities that are based on four equivalence classes of states (panels \textbf{a}-\textbf{i}).
The landscape of $p_\mathrm{AllD}(\mathrm{TFT})$ for different rates of growth ($b-d$) and mutation ($\mu$) can be reconstructed with good agreement (panels \textbf{j}-\textbf{n} and panel \textbf{o}).
The initial population size is $N_0 = 128$. $t_\mathrm{asy}$ and $t_\mathrm{lock}$ are derived from the critical population sizes identified in Fig.~\ref{fig:3}.
}
\label{fig:cond-prop-TFT}
\end{figure}

\newpage
\begin{figure}[h!]
\includegraphics[width=0.9\textwidth, trim=0 0 0 0, clip]{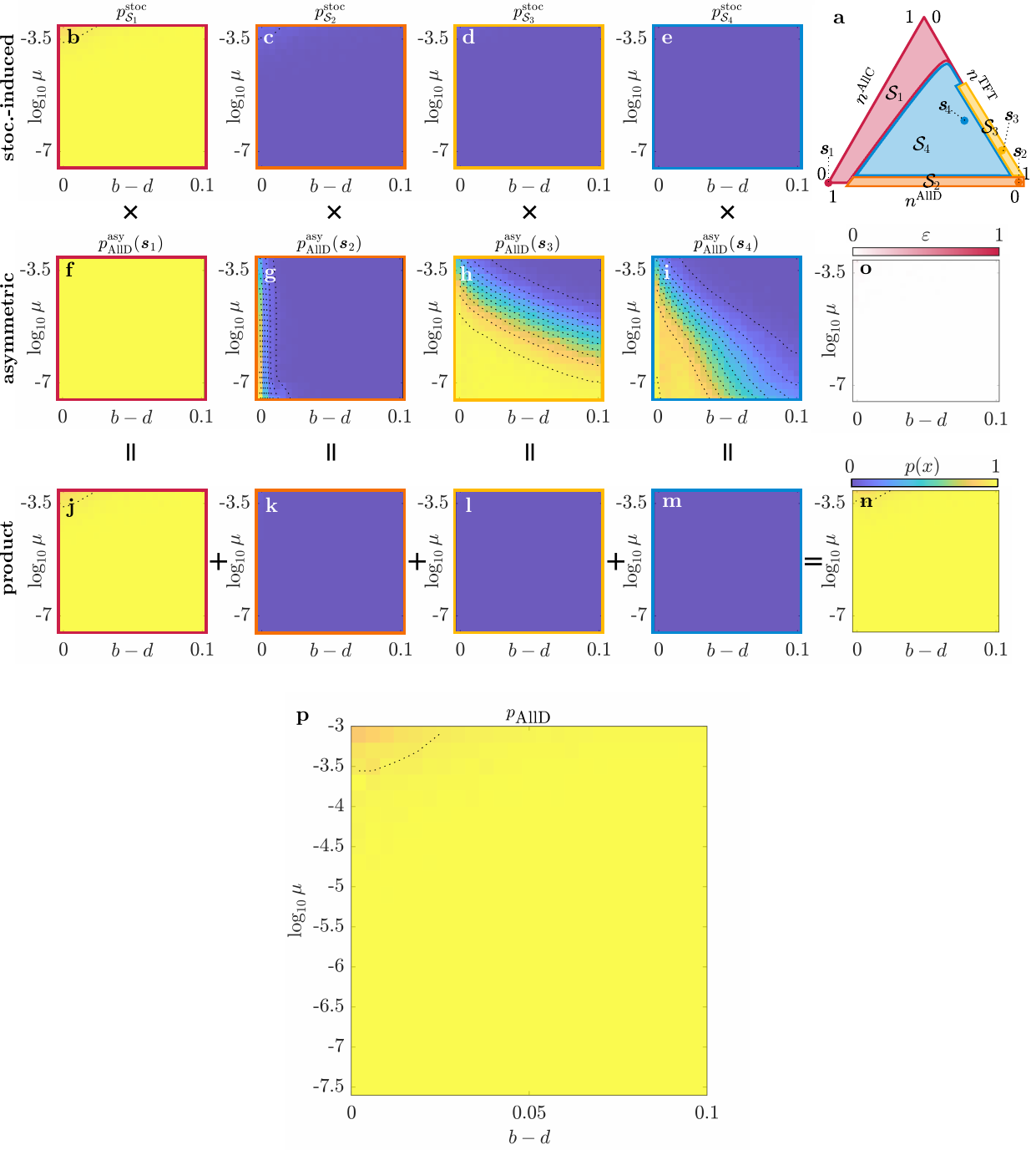}%
\caption{\textbf{Decomposition of outcome statistics based on equivalence classes ($\boldsymbol{n_0=\textrm{AllD-c}}$).}
The statistics of evolutionary outcomes can be decomposed in terms of conditional probabilities that are based on four equivalence classes of states (panels \textbf{a}-\textbf{i}).
The landscape of $p_\mathrm{AllD}(\textrm{AllD-c})$ for different rates of growth ($b-d$) and mutation ($\mu$) can be reconstructed with good agreement (panels \textbf{j}-\textbf{n} and panel \textbf{o}).
The initial population size is $N_0 = 128$. $t_\mathrm{asy}$ and $t_\mathrm{lock}$ are derived from the critical population sizes identified in Fig.~\ref{fig:3}.
}
\label{fig:cond-prop-AllD}
\end{figure}


\newpage
\subsection{Outcome statistics over initial population's size and mix}
\label{sec:additional-no-mix}

\begin{figure}[b!]
\includegraphics[width=0.89\textwidth, trim=0 0 0 0, clip]{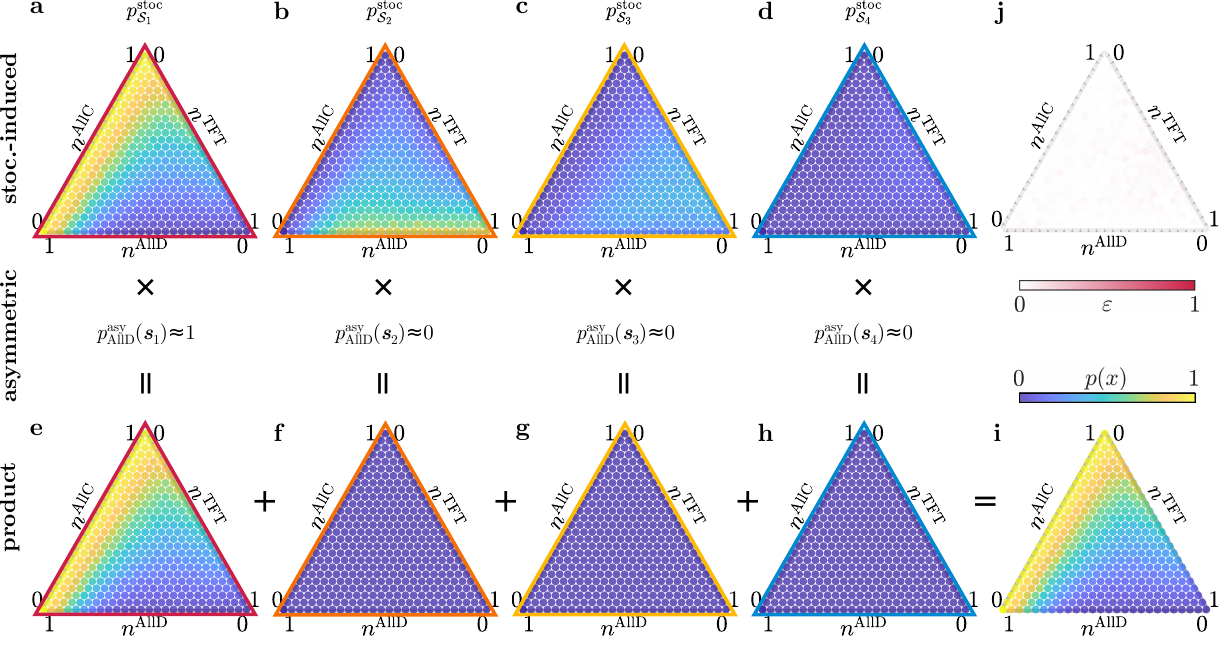}%
\caption{\textbf{Outcome statistics decomposition over $\boldsymbol{n_0}$: \textbf{log}$\boldsymbol{_{10}\mu = -3.5}$, $\boldsymbol{N_0 = 50}$ and $\boldsymbol{b-d=0.05}$.}
Panels \textbf{a-d} show the probability of the system being in the four equivalence classes ${\cal S}_i$ at the end of the stochastically-induced regime.
The probabilities of AllD outcomes given that the system is in the representative points $\boldsymbol{s}_i$ at time beginning of the asymmetric regime are reported in the second row.
Panels \textbf{e-i} show how the probabilities are combined to approximate the outcome probability and panel \textbf{j} shows accuracy of such approximation (cf. Fig.~\ref{fig:2}\textbf{b}).}
\label{fig:dec-high-mu-low-pop}
\end{figure}

In this section we illustrate how the outcome statistics in Fig.~\ref{fig:2} panels \textbf{b}, \textbf{c}, \textbf{d}, \textbf{e} and \textbf{g} can be reconstructed using our probability decomposition method.

At high mutation rate (Figs.~\ref{fig:2}\textbf{b-c}), the asymmetric phase is very short.
This is reflected in the probability decomposition in Figs.~\ref{fig:dec-high-mu-low-pop} and \ref{fig:dec-high-mu-high-pop}, where the probabilities in panels \textbf{f-h} are all zeros, i.e., the statistics of the outcome are given exclusively by the fixation of the system into one of the two basins during the stochastically induced phase (panels \textbf{a}).
The accuracy of the probability decomposition is very high for both small (Fig.~\ref{fig:dec-high-mu-low-pop}) and large (Fig.~\ref{fig:dec-high-mu-high-pop}) initial populations, as illustrated in panels \textbf{j}.

At low mutation rate the asymmetric regime is very long, however, such regime has little effect on the outcome statistics for small initial populations.
This is explained by the probability decomposition in Fig.~\ref{fig:dec-low-mu-low-pop}.
For some starting points $\boldsymbol{N}_0$ the system can be in the equivalence class ${\cal S}_2$ at the end of the stochastically-induced regime (panel \textbf{b}), however, from this area of the simplex it is extremely unlikely for the system to cross the separatrix towards the AllD basin, even if the asymmetric regime is long (panel \textbf{f}).
Moreover, it very unlikely for the system to be in ${\cal S}_4$ at the of the stochastically induced regime, since with a small initial population such regime is longer and fluctuations are likely to drive the systems towards the edges and corners (panels \textbf{d}).
Thus, even if the system can cross the separatrix from ${\cal S}_4$ during the asymmetric regime, the combined probability in panel \textbf{h} is approximately zero for every initial condition.
Finally, we can see that for some initial state $s$ the system has a small chance of being in ${\cal S}_3$ at the end of the stochastically-induced regime (panel \textbf{c}).
From this area the system is very likely to cross the separatrix during the asymmetric regime, leading to the small outcome probability contribution in panel \textbf{g}.

Since the probabilities in panel \textbf{h} are very low, the accuracy of the probability decomposition is very high (panel \textbf{j}) also in this case.
The accuracy gets worse only for middle values of growth rate (around $b-d=0.05$), very small mutation rates (e.g., $\log_{10}\mu<-6$) and large initial populations (e.g., $N_0>200$).
This case is reported in Fig.~\ref{fig:6}, which shows that the probability of AllD outcomes is underestimated for some initial starts around the mixed-strategy unstable fixed point.

Fig.~\ref{fig:2} panels \textbf{d} and \textbf{g} can be reconstructed in a similar way, as shown in Figs.~\ref{fig:dec-fix-N0-high-mu} and \ref{fig:dec-fix-N0-low-mu}.

\begin{figure}[t!]
\includegraphics[width=0.89\textwidth, trim=0 0 0 0, clip]{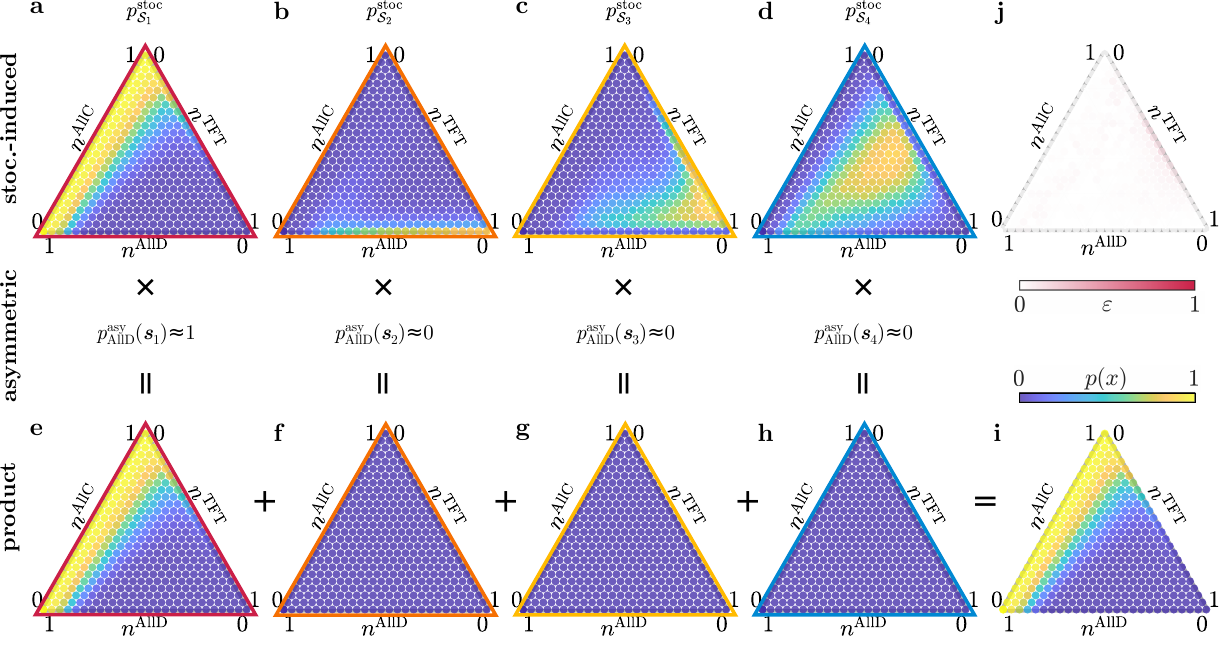}%
\caption{\textbf{Outcome statistics decomposition over $\boldsymbol{n_0}$: \textbf{log}$\boldsymbol{_{10}\mu = -3.5}$, $\boldsymbol{N_0 = 250}$ and $\boldsymbol{b-d=0.05}$.}
Panels \textbf{a-d} show the probability of the system being in the four equivalence classes ${\cal S}_i$ at the end of the stochastically-induced regime.
The probabilities of AllD outcomes given that the system is in the representative points $\boldsymbol{s}_i$ at time beginning of the asymmetric regime are reported in the second row.
Panels \textbf{e-i} show how the probabilities are combined to approximate the outcome probability and panel \textbf{j} shows accuracy of such approximation (cf. Fig.~\ref{fig:2}\textbf{c}).}
\label{fig:dec-high-mu-high-pop}
\end{figure}

\begin{figure}[h!]
\includegraphics[width=0.89\textwidth, trim=0 0 0 0, clip]{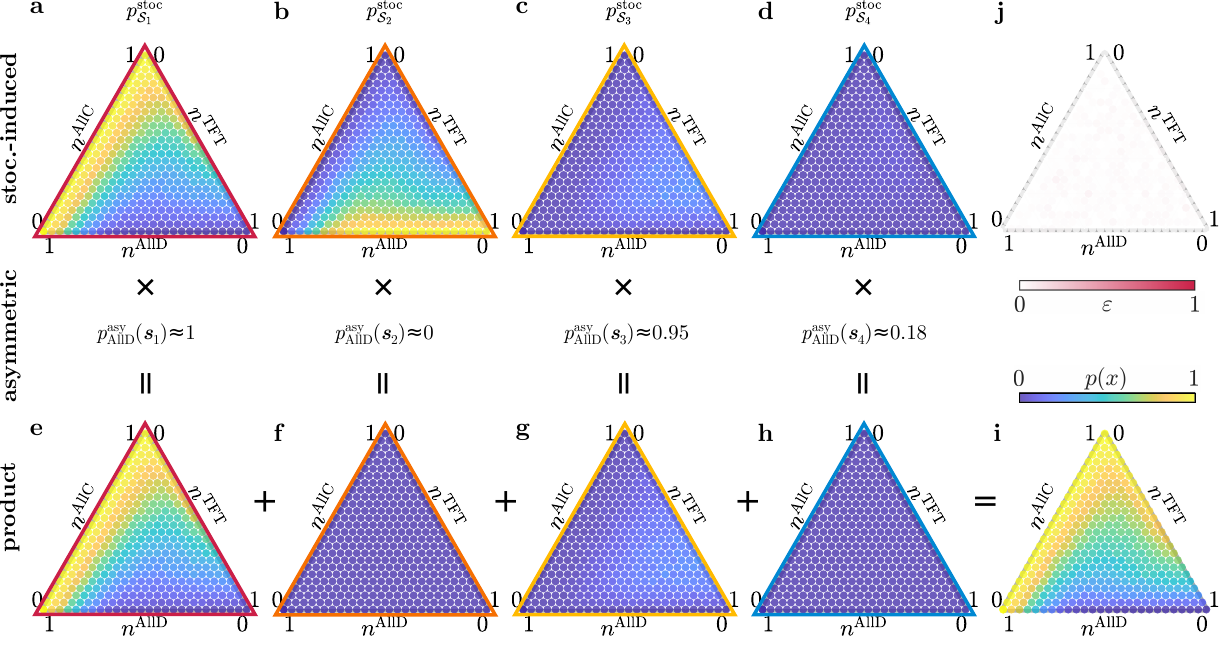}%
\caption{\textbf{Outcome statistics decomposition over $\boldsymbol{n_0}$: \textbf{log}$\boldsymbol{_{10}\mu = -6.5}$, $\boldsymbol{N_0 = 50}$ and $\boldsymbol{b-d=0.05}$.}
Panels \textbf{a-d} show the probability of the system being in the four equivalence classes ${\cal S}_i$ at the end of the stochastically-induced regime.
The probabilities of AllD outcomes given that the system is in the representative points $\boldsymbol{s}_i$ at time beginning of the asymmetric regime are reported in the second row.
Panels \textbf{e-i} show how the probabilities are combined to approximate the outcome probability and panel \textbf{j} shows accuracy of such approximation (cf. Fig.~\ref{fig:2}\textbf{e}).}
\label{fig:dec-low-mu-low-pop}
\end{figure}

\begin{figure}[h!]
\includegraphics[width=0.97\textwidth, trim=0 0 0 0, clip]{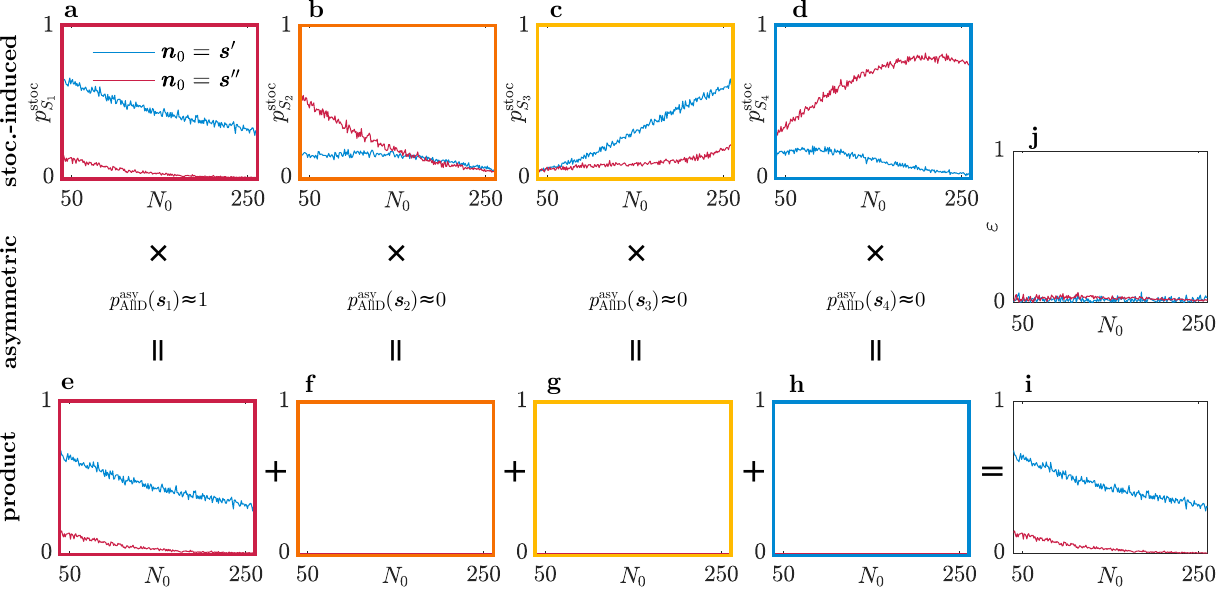}%
\caption{\textbf{Outcome statistics decomposition over $\boldsymbol{N_0}$: \textbf{log}$\boldsymbol{_{10}\mu = -3.5}$ and $\boldsymbol{b-d=0.05}$.}
The results are shown for the two initial states $\boldsymbol{s'}$ and $\boldsymbol{s''}$ in Fig.~\ref{fig:2}.
Panels \textbf{a-d} show the probability of the system being in the four equivalence classes ${\cal S}_i$ at the end of the stochastically-induced regime.
The probabilities of AllD outcomes given that the system is in the representative points $\boldsymbol{s}_i$ at time beginning of the asymmetric regime are reported in the second row.
Panels \textbf{e-i} show how the probabilities are combined to approximate the outcome probability and panel \textbf{j} shows accuracy of such approximation (cf. Fig.~\ref{fig:2}\textbf{d}).}
\label{fig:dec-fix-N0-high-mu}
\end{figure}

\begin{figure}[h!]
\includegraphics[width=0.97\textwidth, trim=0 0 0 0, clip]{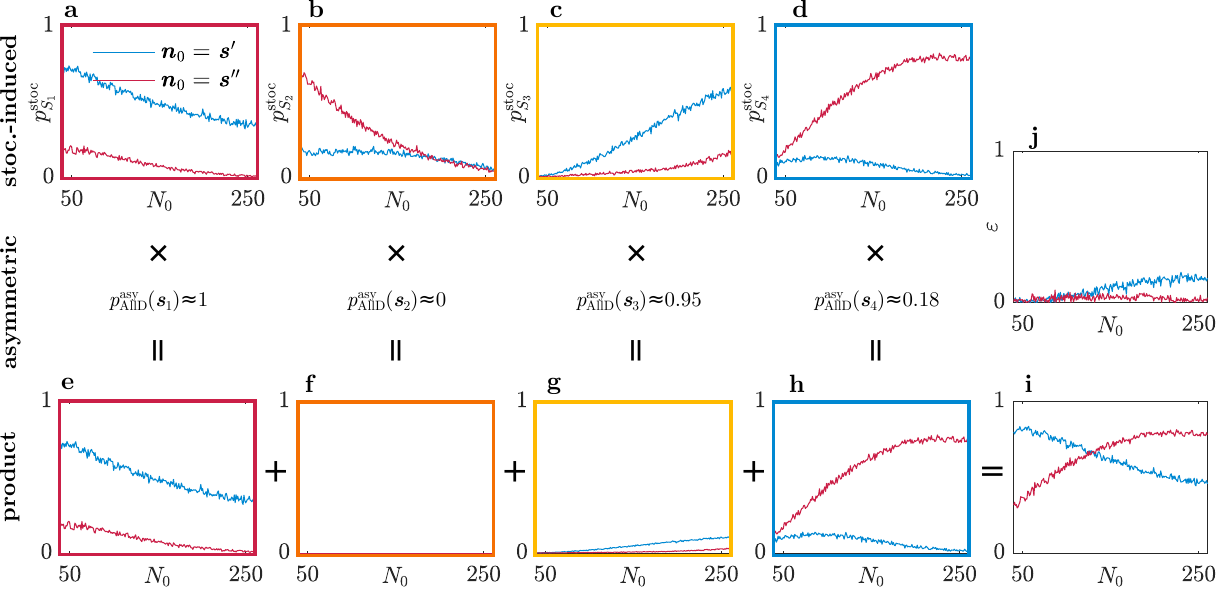}%
\caption{\textbf{Outcome statistics decomposition over $\boldsymbol{N_0}$: \textbf{log}$\boldsymbol{_{10}\mu = -6.5}$ and $\boldsymbol{b-d=0.05}$.}
The results are shown for the two initial states $\boldsymbol{s'}$ and $\boldsymbol{s''}$ in Fig.~\ref{fig:2}.
Panels \textbf{a-d} show the probability of the system being in the four equivalence classes ${\cal S}_i$ at the end of the stochastically-induced regime.
The probabilities of AllD outcomes given that the system is in the representative points $\boldsymbol{s}_i$ at time beginning of the asymmetric regime are reported in the second row.
Panels \textbf{e-i} show how the probabilities are combined to approximate the outcome probability and panel \textbf{j} shows accuracy of such approximation (cf. Fig.~\ref{fig:2}\textbf{g}).}
\label{fig:dec-fix-N0-low-mu}
\end{figure}

\end{widetext}

\end{document}